\documentclass[letterpaper,titlepage,11pt]{article}
\pdfoutput=1
\usepackage[colorlinks=true,urlcolor=blue,citecolor=magenta]{hyperref}
\usepackage{amssymb,amsmath,amsfonts}
\usepackage{epsfig}
\usepackage{epsfig}
\usepackage{graphicx}
\usepackage{epstopdf}
\usepackage{caption}
\usepackage{amscd}
\usepackage{amsthm}
\usepackage{latexsym}
\usepackage{amsbsy}
\usepackage{bbm}
\usepackage[english]{babel}
\usepackage{psfrag}
\usepackage{tabularx}
\allowdisplaybreaks

\def\clock{{\count0=\time
           \divide\count0 60
           \ifnum\count0<10 0\fi\the\count0
           \multiply\count0 -60 \advance\count0 \time
           :\ifnum\count0<10 0\fi \the\count0
         }}
\newcommand{\timestamp}{{\small\vbox{\hbox{\tt\jobname.tex}
\hbox{\the\day/\the\month/\the\year, \clock}}}}

\newcommand{\be}{\begin{equation}}
\newcommand{\ee}{\end{equation}}
\newcommand{\bea}{\begin{eqnarray}}
\newcommand{\eea}{\end{eqnarray}}

\def\6{\partial} \def\7{\tilde} \def\8{\hat}

\def\d{\dot}

\def\r2{{r^2}}

\def\={{\;=\;}}\def\+{{\;+\;}}

\def\dif{{\rm d}}
\def\deriv{\@ifnextchar[{\@deriv}{\@deriv[]}}
   \def\@deriv[#1]#2#3{\mathchoice%
{{\dif^{#1}#2\over\dif{#3}^{#1}}}{{\dif^{#1}#2/\dif{#3}^{#1}}}
{{\dif^{#1}#2\over\dif{#3}^{#1}}}{{\dif^{#1}#2/\dif{#3}^{#1}}}}

\usepackage{color}

\def\d2derpar#1#2{\mathchoice%
{{\partial^2 #1\over\partial #2^2}}%
{{\partial^2 #1/\partial #2^2}}%
{{\partial^2 #1\over\partial #2^2}}%
{{\partial^2 #1/\partial #2^2}}%
}

\def\dif{{\rm d}}
\def\deriv{\@ifnextchar[{\@deriv}{\@deriv[]}}
   \def\@deriv[#1]#2#3{\mathchoice%
{{\dif^{#1}#2\over\dif{#3}^{#1}}}{{\dif^{#1}#2/\dif{#3}^{#1}}}%
{{\dif^{#1}#2\over\dif{#3}^{#1}}}{{\dif^{#1}#2/\dif{#3}^{#1}}}}

\numberwithin{equation}{section}

\begin{document}

\begin{titlepage}
\rightline{\vbox{   \phantom{ghost} }}

 \vskip 1.8 cm

\centerline{\huge \bf
Torsional Newton--Cartan Geometry and the }
\vskip .3cm

\centerline{\huge \bf Schr\"odinger Algebra}

\vskip 1.5cm

\centerline{\large {{\bf Eric A. Bergshoeff$^1$, Jelle Hartong$^2$, Jan Rosseel$^3$}}}

\vskip 1.0cm

\begin{center}

\sl $^1$Van Swinderen Institute for Particle Physics and Gravity, University of Groningen, \\  Nijenborgh 4, 9747 AG Groningen, The Netherlands.\\
\sl $^2$ The Niels Bohr Institute, Copenhagen University, \\
Blegdamsvej 17, DK-2100 Copenhagen {\O}, Denmark.\\
\sl $^3$ Institute for Theoretical Physics, Vienna University of Technology, \\ Wiedner Hauptstr. 8Ð10/136, A-1040 Vienna, Austria.\\
\vskip 0.4cm

\end{center}
\vskip 0.6cm

\centerline{\small\tt emails: e.a.bergshoeff@rug.nl, hartong@nbi.dk, rosseelj@hep.itp.tuwien.ac.at}

\vskip 1.3cm \centerline{\bf Abstract} \vskip 0.2cm \noindent

We show that by gauging the Schr\"odinger algebra with critical exponent $z$ and imposing suitable curvature constraints, that make diffeomorphisms equivalent to time and space translations, one obtains a geometric structure known as (twistless) torsional Newton--Cartan geometry (TTNC). This is a version of torsional Newton--Cartan geometry (TNC) in which the timelike vielbein $\tau_\mu$ must be hypersurface orthogonal. For $z=2$ this version of TTNC geometry is very closely related to the one appearing in holographic duals of $z=2$ Lifshitz space-times based on Einstein gravity coupled to massive vector fields in the bulk. For $z\neq 2$ there is however an extra degree of freedom $b_0$ that does not appear in the holographic setup. We show that the result of the gauging procedure can be extended to include a St\"uckelberg scalar $\chi$ that shifts under the particle number generator of the Schr\"odinger algebra, as well as an extra special conformal symmetry that allows one to gauge away $b_0$. The resulting version of TTNC geometry is the one that appears in the holographic setup. This shows that Schr\"odinger symmetries play a crucial role in holography for Lifshitz space-times and that in fact the entire boundary geometry is dictated by local Schr\"odinger invariance. Finally we show how to extend the formalism to generic torsional Newton--Cartan geometries by relaxing the hypersurface orthogonality condition for the timelike vielbein $\tau_\mu$.

\end{titlepage}

\newcommand{\red}[1]{{\color{red} #1 \color{black}}}
\newcommand{\blue}[1]{{\color{blue} #1 \color{black}}}
\newcommand{\green}[1]{{\color{green} #1 \color{black}}}


\tableofcontents

\section{Introduction}

Newton--Cartan geometry, i.e. the formalism that was used by Cartan \cite{Cartan1,Cartan2} to give a geometrical description of Newtonian gravity in the spirit of General Relativity, has received renewed attention recently\footnote{We refer to \cite{Dautcourt,Eisenhart,Trautman,Kuenzle:1972zw,Duval:1984cj,Duval:1990hj,Julia:1994bs} for earlier work on Newton--Cartan geometry.}. This interest derives from two developments: the work \cite{Son:2013rqa} which showed the usefulness of Newton--Cartan geometry in the effective field theory description of the quantum Hall effect and the work \cite{Christensen:2013lma,Christensen:2013rfa}, building forth on earlier work \cite{Balasubramanian:2010uk,Donos:2010tu,Cassani:2011sv,Chemissany:2011mb} and \cite{Ross:2009ar,Ross:2011gu,Baggio:2011cp,Mann:2011hg,Griffin:2011xs,Chemissany:2012du}, which showed that the boundary geometry of a specific class of asymptotically locally $z=2$ Lifshitz space-times is described by torsional Newton--Cartan (TNC) geometry. In \cite{Christensen:2013lma,Christensen:2013rfa} this formalism was applied to identify which boundary values of bulk fields act as sources in the dual field theory partition function and to calculate quantities like the boundary energy-momentum tensor and exhibit its Ward identities. The holographic studies are building forth on attempts to extend the AdS/CFT correspondence to gravitational theories that can describe field theories that exhibit non-relativistic scale invariance (see \cite{Balasubramanian:2008dm,Son:2008ye,Kachru:2008yh,Taylor:2008tg} and \cite{McGreevy:2009xe,Hartnoll:2009sz,Sachdev:2010ch} for a review).


Recently the results of \cite{Christensen:2013lma,Christensen:2013rfa} have been generalized showing that TNC geometry appears generically in Lifshitz holography \cite{Hartong:2014oma} (see also \cite{Hartong:2014} for additional results and details concerning \cite{Hartong:2014oma}). Around the same time the works \cite{Banerjee:2014pya,Geracie:2014nka,Gromov:2014vla,Bradlyn:2014wla,Banerjee:2014nja,Brauner:2014jaa,Moroz:2014ska,Geracie:2014zha,Jensen:2014aia,Hartong:2014pma} appeared in which field theories coupled to TNC geometries are studied\footnote{It would be interesting to examine whether TNC geometry can also be seen in the formalism of \cite{Chemissany:2014xpa,Chemissany:2014xsa}.}.

In this paper, we will be mostly concerned with the appearance of Newton--Cartan geometry as it arises in the context of holography on asymptotically Lifshitz space-times which in \cite{Hartong:2014oma,Hartong:2014pma} has been shown to exhibit Schr\"odinger invariance\footnote{For a discussion of non-relativistic conformal symmetries and Newton-Cartan structures, see
\cite{Duval:2009vt}.}. This work forms an essential part of this claim by showing that the boundary geometry can be entirely understood in terms of local Schr\"odinger-type symmetries. Lifshitz space-times have the Lifshitz group consisting of space-time translations, spatial rotations and anisotropic scale transformations as isometry group. How they nevertheless can lead to Schr\"odinger invariance is explained in \cite{Hartong:2014pma} and crucially relies on the TNC boundary geometry. 

Lifshitz space-times are non-relativistic in the sense that the causal structure of the boundary becomes non-relativistic, allowing for a notion of absolute time and space. In order to establish a holographic dictionary that can be used to calculate e.g. correlation functions, a geometric description of these non-relativistic boundaries that is covariant with respect to diffeomorphisms and that emphasizes local symmetries, is  required. This is precisely what Newton--Cartan geometry and its generalization to TNC geometry can achieve and where it shows its usefulness. 
Further the use of TNC geometry (with the emphasis on torsion) is also crucial in being able to compute quantities such as the energy density and energy flux. 

Newton--Cartan geometry is often discussed in a metric formalism. For practical applications, particularly in Lifshitz holography, a vielbein formalism is useful, as it emphasizes local symmetries and makes them manifest. Indeed, local symmetries can be essential to discuss e.g. the coupling of boundary field theories to Newton--Cartan geometry. It was recently shown in \cite{Andringa:2010it,Andringa:2013zja} how a vielbein formalism for the torsionless case can be obtained via a gauging procedure of the Bargmann algebra, that is the central extension of the Galilei algebra of non-relativistic space-time transformations. In the gauging, one introduces gauge fields for every generator of the algebra. Their transformations and gauge covariant field strengths are determined by the structure constants of the Bargmann algebra. One also imposes constraints on the field strengths, whose aim is to turn some gauge fields into dependent ones and to identify infinitesimal diffeomorphisms on the remaining independent gauge fields as local space-time translations. In this way, one obtains independent gauge fields $\tau_\mu$, $e_\mu{}^a$ for time and spatial translations resp., that play the role of vielbeine for Newton--Cartan geometry, as well as a gauge field $m_\mu$ for the central charge transformation. Let us emphasize that the inclusion of the latter is crucial for the description of Newton--Cartan geometry. It plays an essential role in turning the gauge fields for spatial rotations and non-relativistic boosts into dependent ones. These dependent gauge fields play a similar role as the spin connection in relativistic gravity and can be used to extract the metric formalism from the vielbein formalism. In particular, an affine connection can be defined via the imposition of a vielbein postulate. In this way, it was shown that the gauging of the Bargmann algebra leads to a description of torsionless Newton--Cartan geometry. The absence of torsion implies that the temporal vielbein $\tau_\mu$ corresponds to a closed one-form and that it can be used to define an absolute time in the space-time.

The aim of this paper is to show how torsional Newton--Cartan geometry in Lifshitz holography can be described in similar terms, i.e. in a vielbein formalism and emphasizing which local symmetries are present and how they are realized. To determine which local symmetries one should look at, we notice that in Lifshitz holography the boundary data are only determined up to anisotropic scale transformations. This leads one to consider non-relativistic conformal algebras. Since the central charge gauge field $m_\mu$ plays a crucial role in obtaining torsionless Newton--Cartan geometry from the Bargmann algebra and moreover played a prominent role in \cite{Christensen:2013lma,Christensen:2013rfa}, we are led to look at the conformal extension of the Bargmann algebra, which is known as the Schr\"odinger algebra. In the first part of this paper, we will therefore show how the appearance of torsion in Newton--Cartan geometry can be  understood from the perspective of gauging the Schr\"odinger algebra. We will perform this gauging for generic dynamical exponent $z$ \footnote{For $z\neq2$, the gauge transformation associated to $m_\mu$ is no longer central. With abuse of terminology, we will however continue to refer to it as the central charge transformation.} and we will argue that it leads to a specific type of TNC geometry, that was dubbed twistless torsional Newton--Cartan geometry (TTNC) in \cite{Christensen:2013lma,Christensen:2013rfa}. TTNC geometry is characterized by the fact that the temporal vielbein is hypersurface orthogonal but not necessarily closed. 

The formulation of TNC geometry that we obtain in the first part of this paper resembles the one that appeared in Lifshitz holography, but also differs from it in a number of respects. In particular, we only obtain TTNC geometry, while in Lifshitz holography more general torsion is allowed. Moreover, we will find some peculiarities, with respect to \cite{Christensen:2013lma,Christensen:2013rfa}. In particular, the torsionful affine connection we are led to in this first part still transforms under central charge transformations, unlike what is found in holographic applications. In the $z\neq2$ case, we will also find that the simple gauging of the Schr\"odinger algebra leads to an extra field, not present in the description of TNC geometry of \cite{Christensen:2013lma,Christensen:2013rfa}. These peculiarities can ultimately be attributed to a difference in how the central charge appears. In the gauging of the first part of this paper, the central charge gauge field will be associated to an abelian central charge gauge symmetry that can be used to remove one of its components. In the description of TNC geometry found in \cite{Christensen:2013lma,Christensen:2013rfa} the central charge is however promoted to a St\"uckelberg symmetry, in the sense that it is accompanied by an extra scalar that shifts under the central charge. An extra component is thus introduced in the formalism.

In a second part of this paper, we will therefore consider how the results of the first part can be extended to deal with this difference, i.e. how the scalar that shifts under the central charge can be incorporated. Since a scalar is not a gauge field, this is no longer an algebra gauging in the strict sense. Nevertheless, we will show how transformation rules and gauge covariant curvatures for gauge fields corresponding to Schr\"odinger-type symmetries can be defined, in the presence of this extra scalar. Curvature constraints can then be imposed that turn some of the gauge fields into dependent ones. The remaining independent fields then correspond to the ones describing TNC geometry, as it appeared in \cite{Christensen:2013lma,Christensen:2013rfa}. This is so even for $z\neq2$, due to the possibility of adding an extra symmetry. We will also show that the dependent gauge fields lead, via a vielbein postulate, to a torsionful connection, that is inert under central charge transformations, as in holographic applications. We will do this analysis first for TTNC geometry, but we will also show how it can be extended to general TNC geometry.

The outline of this paper is as follows. In section \ref{sec:sec2}, we describe the gauging of the Schr\"odinger algebra. After an outline of the Schr\"odinger algebra, we describe the gauging in detail for the $z=2$ case, paying special attention to the appearance of TTNC geometry. We will also show how by choosing a special gauge fixing and reference frame, the result can be expressed in terms of a single Newton potential, similar to what happened in the Bargmann case \cite{Andringa:2010it,Andringa:2013zja}. We will apply a similar procedure for the $z\neq2$ case, pointing out the appearance of an extra field, that is not present in holographic applications. In section \ref{sec:sec3}, we will describe how the results of section \ref{sec:sec2} can be generalized to the case in which the central charge is promoted to a St\"uckelberg symmetry, via the inclusion of a scalar that shifts under the central charge. We will do this for the case of TTNC geometry, both for $z=2$ and $z\neq2$. In the latter case, special attention will be devoted to an extra symmetry that enables one to gauge fix the extra field that appeared in the gauging of section \ref{sec:sec2}. Finally, in section \ref{sec:TNC} the results of section \ref{sec:sec3} will be extended to generic torsion. We conclude in section \ref{sec:conclusions}.

While this paper is mostly concerned with the technical link between Schr\"odinger-type symmetries and torsional Newton--Cartan geometry, the results are expected to have important consequences for Lifshitz holography and non-relativistic field theory on TNC backgrounds, some of which are worked out in \cite{Hartong:2014oma,Hartong:2014,Hartong:2014pma}.

\section{Gauging the Schr\"odinger algebra} \label{sec:sec2}

\subsection{The Schr\"odinger algebra}

The Schr\"odinger algebra is a conformal extension of the Bargmann algebra, the central extension of the Galilei algebra of non-relativistic space-time transformations. In particular, the Schr\"odinger algebra contains a dilatation generator $D$ that acts as an anisotropic scale transformation on the time coordinate $t$ and the $d$ spatial coordinates $x^a$:
\begin{equation}\label{scaletransf}
t \rightarrow \lambda^z t \,, \qquad x^a \rightarrow \lambda x^a \,, \quad a=1,\cdots,d \,.
\end{equation}
The exponent $z$ is called the dynamical exponent and the Schr\"odinger algebra featuring the above scale transformation will be denoted by $\mathtt{sch}_z(d+3)$.

For $z=2$ the Schr\"odinger algebra $\mathtt{sch}_2(d+3)$ is obtained by adding the above dilatation $D$ as well as a special conformal transformation $K$ to the Bargmann algebra, whose generators we will denote by $H$ (time translation), $P_a$ (spatial translations), $G_a$ (Galilean boosts), $J_{ab}$ (spatial rotations) and $N$ (central charge). The commutation relations of $\mathtt{sch}_2(d+3)$ can be obtained by noting that this algebra can be viewed as a subalgebra of  $\mathtt{so}(d+2,2)$, the Lie algebra of the conformal group of $(d+2)$-dimensional Minkowski space-time. We will denote the generators of the latter algebra by $P_\alpha$ (translations), $K_\alpha$ (relativistic special conformal transformations), $\tilde{D}$ (dilatation), $M_{\alpha \beta}$ (Lorentz transformations), with $\alpha=0,1,\cdots,d+1$. Their non-zero commutation relations are given by (with $\eta_{\alpha \beta}$ the $(d+2)$-dimensional Minkowski metric):
\begin{alignat}{2}\label{eq:relconfalg}
\left[\tilde D\,,P_{\alpha}\right] &= -P_{\alpha}\,, & \qquad \left[\tilde D\,,K_{\alpha}\right] &=  K_{\alpha}\,, \nonumber \\ 
\left[P_\alpha\,, K_\beta\right] &= -2\eta_{\alpha\beta}\tilde D+2M_{\alpha\beta}\,,&  \left[M_{\alpha\beta}\,,P_{\gamma}\right] &= \eta_{\alpha\gamma}P_{\beta}-\eta_{\beta\gamma}P_{\alpha}\,,\nonumber\\
\left[M_{\alpha\beta}\,,K_{\gamma}\right] &=  \eta_{\alpha\gamma}K_{\beta}-\eta_{\beta\gamma}K_{\alpha}\,, \nonumber \\
\left[M_{\alpha\beta}\,,M_{\gamma\delta}\right] &=  \eta_{\alpha\gamma}M_{\beta\delta}-\eta_{\alpha\delta}M_{\beta\gamma} -\eta_{\beta\gamma}M_{\alpha\delta}+\eta_{\beta\delta}M_{\alpha\gamma}\,.	
\end{alignat}
The subalgebra $\mathtt{sch}_2(d+3)$ is then defined by the following identifications
\begin{alignat}{2}\label{eq:embedding}
H &=  \frac12\left(P_0+P_{d+1}\right)\,, & \qquad N  &=  P_0-P_{d+1}\,,\nonumber\\
D &=  M_{0\, d+1}+\tilde D\,,&  K  &=  \frac{1}{2}\left(K_0-K_{d+1}\right)\,,\nonumber\\
G_a & = M_{a\, d+1}-M_{a0}\,,
\end{alignat}
and by restricting $\alpha=0,1,\ldots, d,d+1$ to $a=1,\ldots,d$ to obtain $P_a$ and $J_{ab}$  from $P_{\alpha}$ and $M_{\alpha\beta}$. By using these identifications and the commutation relations \eqref{eq:relconfalg}, the following non-zero commutators of $\mathtt{sch}_2(d+3)$ are obtained:
\begin{alignat}{2} \label{Schroedingeralgebra}
\left[D,H\right] &= -2H\,,& \qquad \left[H,K\right] &=  D\,,\nonumber \\
\left[D,K\right] &= 2K\,,& \left[H,G_a\right] &= P_a\,,\nonumber \\
\left[D,P_a\right] &= -P_a\,, & \left[D,G_a\right] &=  G_a\,,\nonumber \\
\left[K,P_a\right] &= -G_a\,,& \left[P_a,G_b\right] &=  \delta_{ab}N\,,\nonumber \\
\left[J_{ab},P_c\right] &= \delta_{ac}P_b-\delta_{bc}P_a\,,& \left[J_{ab},G_c\right] & = \delta_{ac}G_b-\delta_{bc}G_a\,,\nonumber \\
\left[J_{ab},J_{cd}\right] &= \delta_{ac}J_{bd}-\delta_{ad}J_{bc}
	-\delta_{bc}J_{ad}+\delta_{bd}J_{ac}\,.
\end{alignat}
Note that the central charge $N$ of the Bargmann algebra is still a central element of the $z=2$ Schr\"odinger algebra. Furthermore, the triplet $H,D,K$ forms an $\mathtt{sl}(2,\mathbb{R})$ subalgebra. 

For generic dynamical exponent $z\neq1,2$, the Schr\"odinger algebra $\mathtt{sch}_z(d+3)$ is obtained  by modifying the embedding \eqref{eq:embedding} by excluding the generator $K$ and identifying the dilatation generator as
\begin{equation}
D = (z-1)M_{0\, d+1}+\tilde D\,.
\end{equation}
With respect to \eqref{Schroedingeralgebra}, the following commutators are then changed (apart from excluding commutators involving the special conformal transformation $K$):
\begin{eqnarray}
\left[D,H\right] & = & -zH\,,\nonumber \\
\left[D,N\right] & = & (z-2)N\,,\nonumber\\
\left[D,G_a\right] & = & (z-1)G_a\,.
\end{eqnarray}
Note that the generator $N$ no longer corresponds to a central element for $z\neq2$. We will however, with a slight abuse of terminology, still refer to it as the central charge.

Finally, we note that in case $z=1$, the scale transformation \eqref{scaletransf} is compatible with relativistic Lorentz transformations. Indeed, in that case one finds that the Schr\"odinger algebra can be enhanced to the full conformal group of $(d+1)$-dimensional Minkowski space-time. Since this paper is concerned with the non-relativistic Schr\"odinger algebra, we will  not consider the case $z=1$.

\subsection{Gauging the Schr\"odinger algebra for $z=2$} \label{sec:sec22}

In this section we will perform the gauging of the Schr\"odinger algebra with dynamical exponent $z=2$, along similar lines as was done for the Bargmann algebra in \cite{Andringa:2010it,Andringa:2013zja}. In the latter case, the gauging was shown to give the geometrical structure of torsionless Newton--Cartan geometry. Here, we will show that gauging the $z=2$ Schr\"odinger algebra leads to the inclusion of torsion. In particular, we will show that it leads to a formulation of twistless torsional Newton--Cartan geometry, that is covariant with respect to general coordinate transformations and local Schr\"odinger transformations. We will finally show how, upon performing a partial gauge fixing of these symmetries and restricting to a flat background, the fields that define twistless torsional Newton--Cartan geometry reduce to a single Newton potential.

\subsubsection{Gauge transformations and constraints}\label{subsec:gaugetrafosandconstraints}

{\small
\begin{table}[t]
\begin{center}
\begin{tabular}{|c|c|c|c|c|}
\hline
symmetry&generators& gauge field&parameters&curvatures\\[.1truecm]
\hline\rule[-1mm]{0mm}{6mm}
time translations&$H$&$\tau_\mu$&$\zeta(x^\nu)$&$R_{\mu\nu}(H)$\\[.1truecm]
space translations&$P_a$&$e_\mu{}^a$&$\zeta^a(x^\nu)$&$R_{\mu\nu}{}^a(P)$\\[.1truecm]
boosts&$G_a$&$\omega_\mu{}^{a}$&$\lambda^a(x^\nu)$&$R_{\mu\nu}{}^a(G)$\\[.1truecm]
spatial rotations&$J_{ab}$&$\omega_\mu{}^{ab}$&$\lambda^{ab}(x^\nu)$&$R_{\mu\nu}{}^{ab}(J)$\\[.1truecm]
central charge transf.&$N$&$m_\mu$&$\sigma(x^\nu)$&$R_{\mu\nu}(N)$\\[.1truecm]
dilatations&$D$&$b_\mu$&$\Lambda_D(x^\nu)$&$R_{\mu\nu}(D)$\\[.1truecm]
spec. conf. transf.&$K$&$f_\mu$&$\Lambda_K(x^\nu)$&$R_{\mu\nu}(K)$\\[.1truecm]
\hline
\end{tabular}
\end{center} \caption{Summary of the generators of the Schr\"odinger algebra, their associated gauge fields, local parameters and covariant curvatures.}
\label{table:zis2schr}
 \end{table}
 }

The gauging procedure starts by associating a gauge field and corresponding gauge transformation to every generator of the $z=2$ Schr\"odinger algebra. We have summarized our notation for the gauge fields associated to the various transformations of the Schr\"odinger algebra in table \ref{table:zis2schr}. The transformations $G_a$, $J_{ab}$, $N$, $D$, $K$ (i.e. the Schr\"odinger algebra symmetries excluding space-time translations) will often be referred to as `internal symmetries' in this paper. 
The transformations of the various gauge fields under these internal symmetries can be compactly written as
\begin{equation}\label{eq:YMtrafo}
\delta{\mathcal{A}}_{\mu}=\partial_\mu\Sigma+[{\mathcal{A}}_{\mu}\,,\Sigma]\,,
\end{equation}
where $\mathcal{A}_{\mu}$ and $\Sigma$ are Schr\"odinger Lie algebra-valued and given by
\begin{eqnarray}
\mathcal{A}_\mu & = & H\tau_{\mu}+P_ae_{\mu}^a+G_a\omega_{\mu}{}^a+\frac{1}{2}J_{ab}\omega_{\mu}{}^{ab}+N m_{\mu}+Db_{\mu}+Kf_\mu\,,\label{eq:Schconnection}\\
\Sigma & = & G_a\lambda^a+\frac{1}{2}J_{ab}\lambda^{ab}+N\sigma+D\Lambda_D+K\Lambda_K\,.\label{eq:Sigma}
\end{eqnarray}

Writing out the transformation rule \eqref{eq:YMtrafo}, using the Schr\"odinger algebra \eqref{Schroedingeralgebra} one finds that the transformation rules of the various gauge fields under $G_a$, $J_{ab}$, $N$, $D$, $K$ transformations are given by:
\begin{align} \label{gaugetrafoszis2}
\delta\tau_\mu & =  2\Lambda_D\tau_\mu\,,\nonumber \\
\delta e_\mu{}^a & =  \lambda^a{}_b e_\mu{}^b+\lambda^a\tau_\mu+\Lambda_De_\mu{}^a\,,\nonumber \\
\delta\omega_\mu{}^{ab} & =  \partial_\mu\lambda^{ab}+2\lambda^{c[a}\omega_{\mu}{}^{b]}{}_c\,,\nonumber \\
\delta\omega_\mu{}^a & =  \partial_\mu\lambda^a-\lambda^b\omega_\mu{}^a{}_b+\lambda^{a}{}_b\omega_{\mu}{}^b+\lambda^ab_\mu-\Lambda_D\omega_\mu{}^a+\Lambda_K e_\mu{}^a\,, \nonumber \\
\delta b_\mu & =  \partial_\mu\Lambda_D+\Lambda_K\tau_\mu\,, \nonumber \\
\delta f_\mu & =  \partial_\mu\Lambda_K+2\Lambda_K b_\mu-2\Lambda_Df_\mu\,,\nonumber \\
\delta m_\mu & =  \partial_\mu\sigma+\lambda^a e_{\mu a}\,.
\end{align}
A gauge covariant Yang--Mills curvature $\mathcal{F}_{\mu\nu}$ can be defined in the usual way as
\begin{equation}\label{eq:YMcurvature}
\mathcal{F}_{\mu\nu}=\partial_\mu\mathcal{A}_{\nu}-\partial_\nu\mathcal{A}_{\mu}+[\mathcal{A}_{\mu}\,,\mathcal{A}_{\nu}]\,.
\end{equation}
Expanding this as
\begin{eqnarray}
\mathcal{F}_{\mu\nu} & = & HR_{\mu\nu}(H)+P_{a}R_{\mu\nu}{}^{a}(P)+G_{a}R_{\mu\nu}{}^{a}(G)+\frac{1}{2}J_{ab}R_{\mu\nu}{}^{ab}(J)\nonumber\\
&&+N R_{\mu\nu}(N)+D R_{\mu\nu}(D)\,,\label{eq:expansionF}
\end{eqnarray}
one finds that the component expressions for the curvatures are given by:
\begin{align} \label{curvatureszis2}
R_{\mu\nu}(H) & =  2\partial_{[\mu}\tau_{\nu]}-4b_{[\mu}\tau_{\nu]}\,,\nonumber \\
R_{\mu\nu}{}^a(P) & =  2\partial_{[\mu}e_{\nu]}{}^a-2\omega_{[\mu}{}^{ab}e_{\nu]b}-2\omega_{[\mu}{}^a\tau_{\nu]}-2b_{[\mu}e_{\nu]}{}^a\,, \nonumber \\
R_{\mu\nu}{}^{ab}(J) & =  2\partial_{[\mu}\omega_{\nu]}{}^{ab}-2\omega_{[\mu}{}^{c[a}\omega_{\nu]}{}^{b]}{}_c\,,\nonumber \\
R_{\mu\nu}{}^a(G) & =  2\partial_{[\mu}\omega_{\nu]}{}^a+2\omega_{[\mu}{}^b\omega_{\nu]}{}^a{}_b-2\omega_{[\mu}{}^ab_{\nu]}-2f_{[\mu}e_{\nu]}{}^a\,,\nonumber \\
R_{\mu\nu}(D) & =  2\partial_{[\mu}b_{\nu]}-2f_{[\mu}\tau_{\nu]}\,,\nonumber\\
R_{\mu\nu}(K) & =  2\partial_{[\mu}f_{\nu]}+4b_{[\mu}f_{\nu]}\,,\nonumber\\
R_{\mu\nu}(N) & =  2\partial_{[\mu}m_{\nu]}-2\omega_{[\mu}{}^ae_{\nu]a}\,.
\end{align}
These component curvatures obey Bianchi identities, that can compactly be summarized as:
\begin{equation} \label{compBianchi}
D_{[\mu} \mathcal{F}_{\nu \rho]} = \partial_{[\mu} \mathcal{F}_{\nu \rho]} + \left[\mathcal{A}_{[\mu}, \mathcal{F}_{\nu \rho]} \right] = 0 \,.
\end{equation}

The gauge fields $\tau_\mu$ and $e_\mu{}^a$ have the same transformation rules under spatial rotations and Galilean boosts as in the case of the gauging of the Bargmann algebra. In that case, they were identified with vielbeine for the temporal and spatial metric of Newton--Cartan geometry and a similar identification will hold here. As vielbeine of rank 1 and rank $d$ in a $(d+1)$-dimensional space-time, they are not invertible. Projective inverses $v^\mu$ and $e^\mu{}_a$ can however be defined. We will use the convention of  \cite{Christensen:2013rfa}, for which\footnote{With respect to the notation and convention of \cite{Andringa:2010it}, one has $v^\mu=-\tau^\mu$.}
\begin{alignat}{3}
v^\mu \tau_\mu &= -1\,, & \qquad \qquad  v^\mu e_\mu{}^a &= 0 \,, & \qquad \qquad \tau_\mu e^\mu{}_a &= 0\,,\nonumber \\
e_\mu{}^a e^\mu{}_b &= \delta^a_b\,, & e^\mu{}_a e^a{}_\nu & = \delta^\mu_\nu + v^\mu \tau_\nu \,.
\end{alignat}
The inverse vielbeine $v^\mu$ and $e^\mu{}_a$ can then be used to turn curved $\mu$, $\nu$-indices into flat indices. For instance, for a one-form $X_\mu$, the flat temporal component is given by $X_0\equiv -v^\mu X_\mu$. The flat spatial components are given by $X_a = e^\mu{}_a X_\mu$, where the spatial flat index $a = 1,\cdots,d$. The one-form $X_\mu$ can thus be decomposed as
\begin{equation}
X_\mu= X_0\, \tau_\mu+X_a\, e_{\mu}{}^a\,,
\end{equation}
and similar decompositions can be written down for arbitrary tensors. The flat spatial indices $a$, $b$ can be freely raised and lowered with a Kronecker-delta $\delta^{ab}$ or $\delta_{ab}$. Note that $v^\mu$ and $e^\mu{}_a$ transform as follows under gauge transformations:
\begin{align}\label{eq:trafoinvvielbeins}
\delta v^\mu & =  \lambda^a e^\mu{}_a-2\Lambda_D v^\mu\,,\nonumber \\
\delta e^\mu{}_a & =  \lambda_a{}^b e^\mu{}_b-\Lambda_D e^\mu{}_a\,.
\end{align}

Up to this point, we have merely been writing down a gauge theory of the Schr\"odinger algebra. In particular, all gauge fields introduced are so far interpreted as independent gauge fields and transform under local time and spatial translations. In order to make contact with Newton--Cartan geometry, we would like to interpret some of the gauge fields as dependent ones and we would like to identify the action of infinitesimal diffeomorphisms on the remaining independent gauge fields with the action of local time and spatial translations. In order to achieve these two purposes, we will impose extra constraints on the curvatures \eqref{curvatureszis2}.

We will in particular impose two sets of curvature constraints, namely a first set given by
\begin{alignat}{4}\label{constraintszis2}
R_{\mu \nu}(H) &= 0 \,,  & \qquad R_{\mu \nu}{}^a(P) & = 0 \,,  &  \qquad R_{\mu \nu}(N) &= 0  \,,
\end{alignat}
and a second set given by
\begin{alignat}{2} \label{eq:scalarconstraint}
R_{\mu \nu}(D) & = 0 \,, & \qquad R_{0a}{}^a(G)+2m^bR_{0a}{}^a{}_b(J)+m^bm^cR_{ba}{}^a{}_c(J) & = 0\,.
\end{alignat}
In addition to these, extra constraints can be found by inspection of the Bianchi identities \eqref{compBianchi}, that are obeyed by the gauge covariant curvatures \eqref{curvatureszis2}. The extra constraints that follow from imposing $R_{\mu \nu}(H) = R_{\mu \nu}{}^a(P) = R_{\mu \nu}(N) = R_{\mu \nu}(D)= 0$ in the Bianchi identities are given by:
\begin{alignat}{3}
R_{[abc]}(G)&= 0 \,, & \qquad R_{0[ab]}(G)&= 0 \,, & \qquad R_{ab}{}^c(G)-2R_{0[ab]}{}^c(J)&=0\,,  \nonumber \\
R_{[abc]}{}^d(J)&= 0 \,, & \qquad R_{ab}(K)&=0\,.
\end{alignat}
Let us now discuss the consequences of imposing these constraints. 
The constraints of \eqref{constraintszis2} are the analogues of the constraints imposed in the gauging of the Bargmann algebra \cite{Andringa:2010it,Andringa:2013zja}. Differently from that case however, the constraint $R_{\mu \nu}(H)=0$ can now  be solved for the spatial part of the gauge field $b_\mu$ of dilatations, in terms of $\tau_\mu$ and $v^\mu$:
\begin{equation} \label{solbmu}
b_\mu = \frac{1}{2}v^\nu\left(\partial_\nu\tau_\mu-\partial_\mu\tau_\nu\right) - \left(v^\nu b_\nu\right) \tau_\mu\,,
\end{equation}
where the temporal component $-v^\nu b_\nu$ remains undetermined. The constraints $R_{\mu\nu}{}^a(P)=0$ and $R_{\mu \nu}(N)=0$ can similarly be solved for $\omega_\mu{}^{ab}$ and $\omega_\mu{}^a$, leading to the following solutions for these two connection fields in terms of other gauge fields:
\begin{align}\label{eq:expressionsomegaconnections}
\omega_\mu{}^{ab} & = \frac{1}{2}e^{\nu a}\left(\partial_\nu e_\mu{}^b-\partial_\mu e_\nu{}^b\right)-\frac{1}{2}e^{\nu b}\left(\partial_\nu e_\mu{}^a-\partial_\mu e_\nu{}^a\right)+\frac{1}{2}e_{\mu c}e^{\nu a}e^{\rho b}\left(\partial_\nu e_\rho{}^c-\partial_\rho e_\nu{}^c\right)\nonumber\\
&\quad -\frac{1}{2}e^{\nu a}e^{\rho b}\left(\partial_\nu m_\rho-\partial_\rho m_\nu\right)\tau_\mu+e_\mu{}^a (e^{\nu b} b_\nu )-e_\mu{}^b (e^{\nu a} b_\nu )\,,\nonumber\\
\omega_\mu{}^a & =  \frac{1}{2}v^\nu\left(\partial_\nu e_\mu{}^a-\partial_\mu e_\nu{}^a\right)+\frac{1}{2}v^\nu e^{\rho a}e_{\mu b}\left(\partial_\nu e_\rho{}^b-\partial_\rho e_\nu{}^b\right)+\frac{1}{2}e^{\nu a}\left(\partial_\mu m_\nu-\partial_\nu m_\mu\right)\nonumber\\
&\quad -\frac{1}{2}\tau_\mu v^\rho e^{\nu a}\left(\partial_\rho m_\nu-\partial_\nu m_\rho\right)-\left(v^\nu b_\nu\right) e_\mu{}^a\,.
\end{align}
Note that $m_\mu$ only enters these expressions via its curl $\partial_\mu m_\nu-\partial_\nu m_\mu$. Further one can check that these expressions for $\omega_\mu{}^{ab}$ and $\omega_\mu{}^{a}$ are such that they transform exactly as in \eqref{gaugetrafoszis2} before we imposed the curvature constraints. The constraints in \eqref{eq:scalarconstraint} can be used to solve for $f_\mu$. In particular, $R_{\mu \nu}(D)=0$ can be used to solve for the spatial part of $f_\mu$ in terms of other gauge fields, such that one can write
\begin{equation}
f_\mu = v^\nu \left( \partial_\nu b_\mu - \partial_\mu b_\nu\right) - (v^\nu f_\nu) \tau_\mu \,.
\end{equation}
The temporal part $f_0=-v^\nu f_\nu$ is determined by the second constraint of \eqref{eq:scalarconstraint} and is given by the following expression:
\begin{eqnarray}
v^\mu f_\mu & = & \frac{1}{d}v^\mu e^\nu{}_a\left(\partial_\mu\omega_\nu{}^a-\partial_\nu\omega_\mu{}^a+\omega_\mu{}^b\omega_\nu{}^a{}_b-\omega_\nu{}^b\omega_\mu{}^a{}_b+b_\mu\omega_\nu{}^a-b_\nu\omega_\mu{}^a\right)\nonumber\\
&&+\frac{2}{d}v^\mu m^c R_{\mu a}{}^a{}_c(J)-\frac{1}{d}m^bm^cR_{ba}{}^a{}_c(J)\,.\label{eq:vfz=2}
\end{eqnarray}
The resulting expression for $f_\mu$ transforms as in \eqref{gaugetrafoszis2} under $G_a$, $J_{ab}$, $D$, $K$ transformations. It is however not invariant under $N$ transformations\footnote{Instead of \eqref{eq:scalarconstraint}, we could  have imposed the alternative constraint $R_{0a}{}^a(G)=0$. The resulting solution for $f_\mu$ would not have been boost invariant but it would on the other hand have been invariant under the central charge transformation $N$. In the next subsection we will see another example of a similar kind of tension between having either $G_a$ or $N$ invariance. Since we have already introduced a St\"uckelberg field for boost transformations, i.e. $m_\mu$ we will show in section \ref{sec:Stueckelbergscalar} that by introducing a St\"uckelberg field for $N$ transformations we can make boost invariant expressions also invariant under $N$ gauge transformations. This is why here we choose to impose \eqref{eq:scalarconstraint}.}.

At this point, one is left with the independent fields $\tau_\mu$, $e_\mu{}^a$, $m_\mu$ and $v^\mu b_\mu$. The first three correspond to the independent gauge fields, present in the gauging of the Bargmann algebra. The temporal component $-v^\mu b_\mu$ of the gauge field of dilatations was not present in that case. Using the $b_\mu$ and $v^\mu$ transformation rules given in \eqref{gaugetrafoszis2} and \eqref{eq:trafoinvvielbeins} respectively, one obtains:
\begin{equation}\label{eq:trafovb}
\delta\left(v^\mu b_\mu\right)=v^\mu\partial_\mu\Lambda_D-2\Lambda_Dv^\mu b_\mu+\lambda^a e^\mu{}_ab_\mu-\Lambda_K\,.
\end{equation}
One thus finds that $v^\mu b_\mu$ corresponds to a St\"uckelberg field for special conformal transformations. It can thus easily be chosen to be zero, fixing $K$ transformations. Note that choosing this gauge introduces a compensating $K$ transformation
\begin{equation} \label{Kcompensator}
\Lambda_K = v^\mu \partial_\mu \Lambda_D + \lambda^a e^\mu{}_a b_\mu \,.
\end{equation}
The transformation rules of $\omega_\mu{}^a$ and $b_\mu$ after adopting the gauge fixing condition $v^\mu b_\mu = 0$ are then still as in \eqref{gaugetrafoszis2}, provided that $\Lambda_K$ is interpreted as the compensating transformation \eqref{Kcompensator}. 

Adopting the gauge fixing condition $v^\mu b_\mu = 0$, we are left with independent fields $\tau_\mu$, $e_\mu{}^a$ and $m_\mu$. Having used the constraints to turn some of the gauge fields into dependent ones, let us now discuss the interplay between local space-time translations, infinitesimal diffeomorphisms and constraints. Inspecting the action of an infinitesimal diffeomorphism with parameter $\xi^\mu$ on $\tau_\mu$, we find:
\begin{align}
\delta_\xi \tau_\mu &= \xi^\nu \partial_\nu \tau_\mu + \partial_\mu \xi^\nu \tau_\nu \nonumber \\
& = \partial_\mu \left(\xi^\nu \tau_\nu \right) - 2  \left(\xi^\nu \tau_\nu \right) b_\mu + 2  \left(\xi^\nu b_\nu \right)\tau_\mu - \xi^\nu R_{\mu \nu}(H) \,.
\end{align}
The first two terms correspond to a local $H$ transformation on $\tau_\mu$ with parameter $\xi^\nu \tau_\nu$, while the third term corresponds to a local dilatation with parameter $\xi^\nu b_\nu$ acting on $\tau_\mu$. We thus see that, upon imposing $R_{\mu \nu}(H) = 0$, the action of infinitesimal diffeomorphisms on $\tau_\mu$ can be identified with the action of local $H$ transformations and local dilatations. In a similar manner, one can see that the actions of infinitesimal diffeomorphisms on $e_\mu{}^a$ and $m_\mu$ can be identified with local $H$ and $P_a$ transformations, along with local $G_a$, $J_{ab}$, $N$ and $D$ transformations, once the constraints $R_{\mu \nu}{}^a(P) = R_{\mu \nu}(N) = 0$ are imposed. The action of infinitesimal diffeomorphisms on the independent gauge fields thus agrees with the transformations of the Schr\"odinger algebra. Note that in the explicit expressions for the dependent gauge fields, all independent fields appear either without derivatives or via the curl of a field that transforms properly under diffeomorphisms. The transformation of the dependent gauge fields under diffeomorphisms, that is induced by the one of the independent fields, is thus the expected one.

\subsubsection{Twistless torsional Newton--Cartan geometry}\label{subsec:affineconnection}

The end results of the above gauging of the $z=2$ Schr\"odinger algebra, given in terms of the independent fields $\tau_\mu$, $e_\mu{}^a$ and $m_\mu$, can be interpreted as giving the geometrical data defining a Newton--Cartan geometry. This is similar to the case of the gauging of the Bargmann algebra. In particular, as mentioned above, the gauge fields $\tau_\mu$ and $e_\mu{}^a$ can be viewed as vielbeine for the temporal and spatial Newton--Cartan metrics $\tau_{\mu \nu}$ and $h^{\mu \nu}$
\begin{equation}
\tau_{\mu \nu} = \tau_\mu \tau_\nu \,, \qquad h^{\mu\nu} = e^\mu{}_a \delta^{ab} e^\nu{}_b \,,
\end{equation}
where we have defined the spatial Newton--Cartan metric $h^{\mu \nu}$ with upper indices in terms of the projective inverse vielbein $e^\mu{}_a$. Similarly, we can define a projective inverse $h_{\mu \nu}$ with lower indices as $h_{\mu \nu} = e_\mu{}^a \delta_{ab} e_\nu{}^b$.

As in the Bargmann case, the inclusion of the central charge gauge field $m_\mu$ is crucial to uniquely determine the gauge fields $\omega_\mu{}^{ab}$ and $\omega_\mu{}^a$ in terms of $\tau_\mu$, $e_\mu{}^a$ and $m_\mu$, as solutions of the constraints $R_{\mu \nu}{}^a(P) = 0$ and $R_{\mu \nu}(N) = 0$. These can then be interpreted as spin connections for local spatial rotations and local Galilean boosts respectively and they can be used to define an affine connection $\tilde\Gamma^\rho_{\mu \nu}$ by imposing the vielbein postulates:
\begin{align} \label{vielbposts}
\mathcal{D}_\mu \tau_\nu &\equiv \partial_\mu \tau_\nu - \tilde\Gamma^\rho_{\mu \nu} \tau_\rho -2b_\mu\tau_\nu= 0 \,, \nonumber \\
\mathcal{D}_\mu e_\nu{}^a &\equiv \partial_\mu e_\nu{}^a - \tilde\Gamma^\rho_{\mu \nu} e_\rho{}^a - \omega_\mu{}^a{}_b e_\nu{}^b - \omega_\mu{}^a \tau_\nu-b_\mu e_\nu{}^a = 0 \,.
\end{align}
From the curvature constraints \eqref{constraintszis2} we learn that $\tilde\Gamma^\rho_{\mu \nu}$ is symmetric and thus has no torsion. The connection $\tilde\Gamma^\rho_{\mu \nu}$ is uniquely determined by these vielbein postulates and its explicit expression is given by:
\begin{align}\label{eq:tildeGamma}
\tilde\Gamma^\rho_{\mu \nu} &= -v^\rho \partial_\mu \tau_\nu + \frac12 h^{\rho\sigma} \left(\partial_\mu h_{\sigma \nu} + \partial_\nu h_{\sigma \mu} - \partial_\sigma h_{\mu \nu} \right) + \tau_\mu h^{\rho \sigma} \partial_{[\sigma} m_{\nu]} + \tau_\nu h^{\rho \sigma} \partial_{[\sigma} m_{\mu]} \nonumber \\
& \quad + h^{\rho \sigma} b_\sigma h_{\mu \nu} - \delta_\mu^\rho b_\nu - \delta_\nu^\rho b_\mu -v^\rho( \tau_\mu b_\nu-\tau_\nu b_\mu)\,.
\end{align}
While uniquely determined, the connection $\tilde\Gamma^\rho_{\mu \nu}$ is not metric compatible, where by metric compatibility we mean that
\begin{eqnarray}
\nabla_\mu\tau_\nu & = & 0\,,\label{eq:metriccompatible1}\\
\nabla_\mu h^{\nu\rho} & = & 0\,.\label{eq:metriccompatible2}
\end{eqnarray}
We will however use the connection $\tilde\Gamma^\rho_{\mu\nu}$ to distill the metric compatible connection for torsional Newton--Cartan geometry $\Gamma^\rho_{\mu\nu}$. There are many ways to do this, as one is always free to add a tensor to a connection to obtain a new connection. We will fix this freedom by requiring that the connection $\Gamma^\rho_{\mu\nu}$ is boost invariant. This can be achieved by `throwing away the $b_\mu$ terms' in $\tilde\Gamma^\rho_{\mu \nu}$. We can think of the $b_\mu$ terms as arising from the gauging of dilatations. That is, suppose we are given a metric compatible $\Gamma^\rho_{\mu\nu}$ then $\tilde\Gamma^\rho_{\mu\nu}$ is obtained from $\Gamma^\rho_{\mu\nu}$ by replacing all ordinary derivatives by dilatation covariant derivatives\footnote{Also the way in which the $b_\mu$ terms appear in \eqref{eq:expressionsomegaconnections} is precisely such that all ordinary derivatives can be written as dilatation covariant derivatives.}. Because the $b_\mu$ field is in part a dependent gauge field the procedure by which to drop the $b_\mu$ terms is ambiguous. For example if we drop them in \eqref{eq:tildeGamma}, we get an expression that is invariant under $N$ transformations but not under boosts. The expression \eqref{eq:tildeGamma} after having dropped the $b_\mu$ terms transforms under boosts into terms proportional to $\partial_\mu\tau_\nu-\partial_\nu\tau_\mu$ which via the curvature constraint $R_{\mu\nu}(H)=0$ can be traded for terms containing $b_\mu$. Another way of writing the expression \eqref{eq:tildeGamma} is as follows
\begin{eqnarray}
\tilde\Gamma^\rho_{\mu \nu} &=& -\left(v^\rho-h^{\rho\sigma}m_\sigma\right)\left(\partial_\mu-2b_\mu\right)\tau_\nu+\frac{1}{2}h^{\rho\sigma}\left[\left(\partial_\mu-2b_\mu\right)\left(h_{\nu\sigma}-\tau_\nu m_\sigma-\tau_\sigma m_\nu\right)\right.\nonumber\\
&&\left.+\left(\partial_\nu-2b_\nu\right)\left(h_{\mu\sigma}-\tau_\mu m_\sigma-\tau_\sigma m_\mu\right)-\left(\partial_\sigma-2b_\sigma\right)\left(h_{\mu\nu}-\tau_\mu m_\nu-\tau_\nu m_\mu\right)\right]\,,
\end{eqnarray}
where we have made use of \eqref{solbmu}. This expression is manifestly boost invariant and if we now drop the $b_\mu$ terms we obtain the metric compatible boost invariant connection $\Gamma_{\mu\nu}^\rho$ that is given by
\begin{eqnarray}
\Gamma^\rho_{\mu \nu} &=& -\left(v^\rho-h^{\rho\sigma}m_\sigma\right)\partial_\mu\tau_\nu+\frac{1}{2}h^{\rho\sigma}\left[\partial_\mu\left(h_{\nu\sigma}-\tau_\nu m_\sigma-\tau_\sigma m_\nu\right)\right.\nonumber\\
&&\left.+\partial_\nu\left(h_{\mu\sigma}-\tau_\mu m_\sigma-\tau_\sigma m_\mu\right)-\partial_\sigma\left(h_{\mu\nu}-\tau_\mu m_\nu-\tau_\nu m_\mu\right)\right]\,.\label{eq:Gamma}
\end{eqnarray}
This expression was also found in \cite{Jensen:2014aia}. The torsion is then given by
\begin{equation}
\Gamma^\rho_{[\mu \nu]}=-\frac{1}{2}\left(v^\rho-h^{\rho\sigma}m_\sigma\right)\left(\partial_\mu\tau_\nu-\partial_\nu\tau_\mu\right)\,.
\end{equation}
The fact that adding the $b_\mu$ field to the connection $\Gamma^\rho_{\mu\nu}$ by replacing ordinary derivatives by dilatation covariant ones leads to a connection $\tilde\Gamma^\rho_{\mu\nu}$ that is torsionless tells us that torsion is necessary to make the formalism dilatation covariant. In other words one thus sees that the role of the gauge field of dilatations $b_\mu$ is to introduce torsion in the context of Newton--Cartan geometry.

We note that the expression for $\Gamma^\rho_{\mu\nu}$, given in \eqref{eq:Gamma}, is not invariant under the $N$ gauge transformations. The situation is analogous to what happened when we imposed the curvature constraint to solve for $v^\mu f_\mu$ in the previous section. Either the connection obtained by removing the $b_\mu$ terms from $\tilde\Gamma^\rho_{\mu \nu}$ is $N$ but not boost invariant (see the $b$-independent part of equation \eqref{eq:tildeGamma} which only depends on $m_\mu$ via its curl) or it is boost but not $N$ invariant (see equation \eqref{eq:Gamma}). Again just as in the previous subsection we choose boost over $N$ invariance because we will later introduce a St\"uckelberg field to achieve $N$ invariance whereas we have already done that for boost invariance via the $m_\mu$ gauge connection.

Let us stress the difference with the gauging of the Bargmann algebra. In that case the resulting Newton--Cartan geometry is torsionless; i.e. the affine connection $\Gamma^\rho_{\mu \nu}$ is equal to the standard symmetric Newton--Cartan connection \cite{Andringa:2010it} and hence the temporal vielbein $\tau_\mu$ corresponds to a closed one-form : $\partial_{[\mu} \tau_{\nu]} = 0$. This is no longer the case here. In particular, the constraint $R_{\mu \nu}(H) = 0$ no longer implies that $\tau_\mu$ is closed, but rather that 
\begin{equation} \label{hypsurforth1}
\partial_{[\mu} \tau_{\nu]} = 2 b_{[\mu} \tau_{\nu]} \,.
\end{equation}
The gauge field of dilatations can not be fully gauged away in general; only its temporal component (that does not appear in \eqref{hypsurforth1}) can be gauged away using a special conformal transformation. One thus finds that $\partial_{[\mu}\tau_{\nu]}$ can not be put to zero in general and equation \eqref{hypsurforth1} can be viewed as determining the spatial part of $b_\mu$ in terms of $\partial_{[\mu}\tau_{\nu]}$, as was done in \eqref{solbmu}. Note that \eqref{hypsurforth1}, via Frobenius' theorem, is equivalent to 
\begin{equation} \label{hypsurforth2}
\tau_{[\mu} \partial_\rho \tau_{\nu]} = 0 \,.
\end{equation}
The one-form $\tau_\mu$ is thus hypersurface orthogonal. Physically, this implies that there exists a preferred foliation in equal-time slices for the space-time. These can be thought of as hypersurfaces of absolute simultaneity.

The gauging of the $z=2$ Schr\"odinger algebra leads to a $\tau_\mu$ which according to \eqref{hypsurforth1} must obey the twistlessness condition
\begin{equation}\label{eq:TTNC}
h_\mu{}^\rho h_\nu{}^\sigma\left(\partial_\rho\tau_\sigma-\partial_\sigma\tau_\rho\right)=0\,,
\end{equation}
where $h_\mu{}^\nu=\delta_\mu^\nu+\tau_\mu v^\nu$. The converse is also true, i.e. any solution of the twistlessness condition \eqref{eq:TTNC} is of the form \eqref{hypsurforth1} where $b_\mu$ is thus a field of the form \eqref{solbmu}. The resulting geometry was called twistless torsional Newton--Cartan geometry (TTNC) in \cite{Christensen:2013lma,Christensen:2013rfa}. It is not the most general form of torsional Newton--Cartan geometry that can appear in Lifshitz holography however. It is possible to fully relax $\tau_\mu$ so that it becomes unconstrained. The resulting geometry is an extension of TTNC geometry that is called torsional Newton--Cartan geometry and this will be discussed later in section \ref{sec:TNC}.

\subsubsection{Gauge fixing to acceleration extended Galilei symmetries}\label{subsubsec:gaugefixingz=2}

In the previous section, we have seen that the gauging of the Schr\"odinger algebra in general leads to twistless torsional Newton--Cartan geometry. In the Bargmann case \cite{Andringa:2010it,Andringa:2013zja}, after the gauging of the Bargmann algebra is done, one can perform a partial gauge fixing of all local symmetries to the so-called acceleration extended Galilei symmetries, that extend the Galilei symmetries by allowing for time-dependent spatial translations \cite{Duval:1993pe,Kuchar:1980tw,DePietri:1994je}\footnote{Acceleration extended Galilei symmetries were called `Milne isometries' in \cite{Duval:1993pe}.}. This gauge fixing corresponds to choosing special coordinate frames in which the fields that define Newton--Cartan geometry are determined by a single Newton potential. In this section, we will show that a similar gauge fixing can be performed in case one gauges the Schr\"odinger algebra. Throughout this section, we will split curved $\mu$-indices as $\{0,i\}$, $i=1,\cdots,d$. We will moreover also adopt the convention that parameters of symmetry transformations are assumed to be constant, unless their coordinate dependence is given explicitly.

We will consider the case in which the spatial sections of the Newton--Cartan space-time are flat and we will thus take:
\begin{equation} \label{fixrotcond}
\omega_\mu{}^{ab} = 0 \,.
\end{equation}
This condition fixes the local spatial rotations with parameter $\lambda^{ab}(x^\mu)$ to constant ones, whose parameter will henceforth be denoted by $\lambda^{ab}$.

In order to continue, we note that Frobenius' theorem and the hypersurface orthogonality of $\tau_\mu$ (see eqs. \eqref{hypsurforth1} and \eqref{hypsurforth2}) imply that we can write
\begin{equation}
\tau_\mu = \psi \partial_\mu \tau \,.
\end{equation}
The gauge transformation
\begin{equation}
\delta\tau_\mu =\xi^\nu(x^\rho)\partial_\nu\tau_\mu+\partial_\mu\xi^\nu(x^\rho)\tau_\nu+2\Lambda_D(x^\rho)\tau_\mu\,,
\end{equation}
for $\tau_\mu$ then induces the following transformations for $\psi$ and $\tau$:
\begin{eqnarray}
\delta\psi & = & \xi^\mu(x^\rho)\partial_\mu\psi+2\Lambda_D(x^\rho)\psi\,,\\
\delta\tau & = & \xi^\mu(x^\rho)\partial_\mu\tau + c\,,
\end{eqnarray}
where $c$ represents a constant shift. By setting
\begin{equation}
\psi=1\,,
\end{equation}
we can completely gauge fix the dilatations. By setting $\tau = x^0 = t$, we can gauge fix the $\xi^0(x^\mu)$ transformations to constant ones. We have thus set
\begin{equation} \label{fixtau}
\tau_\mu = \delta_\mu^0 \,,
\end{equation}
leaving us with constant $\xi^0$ transformations and fixing dilatations completely. Noting that then
\begin{equation}
\delta b_0  = \Lambda_K(x^\mu) \,,
\end{equation}
we find that the special conformal $K$ transformation can be completely fixed by putting
\begin{equation}
b_0 = 0 \,.
\end{equation}
From \eqref{solbmu} and \eqref{fixtau}, we then find that $b_\mu = 0$. After these steps, the transformation rules of the remaining independent fields reduce to the ones that appeared in the Bargmann case \cite{Andringa:2013zja}. Similarly, the expressions for the dependent fields $\omega_\mu{}^{ab}$ and $\omega_\mu{}^a$ reduce to the ones of that case. Moreover, the spatial components of $f_\mu$ are zero and the only non-trivial component of $f_\mu$ left is the time-like component $-v^\mu f_\mu$.

From this point on, the gauge fixing procedure of \cite{Andringa:2013zja} can be taken over. In particular, one can set
\begin{equation}\label{eq:choicee}
e_i{}^a = \delta_i{}^a \,,
\end{equation}
to fix $\xi^a(x^\mu)$ to be of the form
\begin{equation} \label{fixedxia}
\xi^a (x^\mu) = \xi^a (t) - \lambda^a{}_b x^b \,.
\end{equation}
Considering the transformation rule for $\tau^a \equiv - e_0{}^a$ under the remaining transformations:
\begin{equation}
\delta \tau^a = \xi^0 \partial_t \tau^a + \xi^i(t) \partial_i \tau^a - \lambda^i{}_j x^j \partial_i \tau^a + \lambda^a{}_b \tau^b - \partial_t \xi^a(t) - \lambda^a(x^\mu) \,,
\end{equation}
one finds that one can use the boosts to put 
\begin{equation}\label{eq:fixboosts}
\tau^a =e_0{}^a= 0\,,
\end{equation}
at the expense of introducing a compensating transformation
\begin{equation} \label{fixedlambda}
\lambda^a(x^\mu) = - \partial_t \xi^a (t) \,.
\end{equation}
Then, only $m_\mu$ is left over as an independent field. Examining the condition $\omega_0{}^{ab} = 0$, leads to the conclusion that
\begin{equation}
\partial_{[i} m_{j]} = 0 \qquad \Longrightarrow \qquad m_i  = \partial_i m \,.
\end{equation}
This leaves us with two fields $m$, $m_0$, whose transformation rules under the remaining transformations are given by
\begin{eqnarray}
\delta m & = & \xi^0 \partial_t m + \xi^i(t) \partial_i m - \lambda^i{}_j x^j \partial_i m - \partial_t \xi^i(t) x^i + \sigma(x^\mu) + Y(t) \,, \label{deltam}\nonumber \\
\delta m_0 & = & \xi^0 \partial_t m_0 + \xi^i(t) \partial_i m_0- \lambda^i{}_j x^j \partial_i m_0 + \partial_t \xi^i(t) \partial_i m + \partial_t \sigma(x^\mu) \,,\label{eq:trafosm0andm}
\end{eqnarray}
where $Y(t)$ is an arbitrary time-dependent shift. One can now use the $\sigma(x^\mu)$ transformation to fix 
\begin{equation} \label{fixm}
m = f(t,x^i) \,,
\end{equation}
where $f(t,x^i)$ is an arbitrary, but fixed function. Imposing that $\delta m = 0$, leads to the following compensating transformation:
\begin{equation}
\sigma(x^\mu) = - \xi^0 \partial_t f - \xi^i(t) \partial_i f + \lambda^i{}_j x^j \partial_i f + \partial_t \xi^i (t) x^i - Y(t) \,.
\end{equation}
The remaining field $m_0$ can then be used to define a field $\Phi$ as:
\begin{equation}\label{eq:deff}
\Phi = m_0 - \partial_t f \,.
\end{equation}
The transformation rule for $\Phi$ is calculated as $\delta \Phi = \delta m_0$, leading to
\begin{equation}\label{eq:trafoNP}
\delta \Phi = \xi^0 \partial_t \Phi + \xi^i(t) \partial_i \Phi - \lambda^i{}_j x^j \partial_i \Phi + \partial_t^2 \xi^i(t) x^i - \partial_t Y(t) \,.
\end{equation}
This is the expected transformation law for the Newton potential under acceleration extended Galilei symmetries and $\Phi$ can thus be identified with the Newton potential. The algebra obeyed by these remaining acceleration extended Galilei symmetries is characterized by the following non-zero commutators:
\begin{align} \label{boscomm}
\left[ \delta_{\xi^{ 0}}, \delta_{\xi^i(t)} \right] \Phi & = \delta_{\xi^i(t)}\left(-\xi^{0} \dot{\xi}^i(t)\right) \Phi \,, \nonumber \\[.1truecm]
\left[ \delta_{\xi^{ 0}}, \delta_{\sigma(t)} \right] \Phi & =  \delta_{\sigma(t)} \left(- \xi^{ 0} \dot{\sigma}(t) \right) \Phi \,, \nonumber \\[.1truecm]
\left[ \delta_{\xi^i_1(t)}, \delta_{\xi^i_2(t)} \right] \Phi & =  \delta_{\sigma(t)}\left(\dot{\xi}_1^j (t) \xi_2^j (t) - \dot{\xi}_2^j (t) \xi_1^j (t)\right) \Phi \,, \nonumber \\[.1truecm]
\left[\delta_{\xi^i(t)} ,\delta_{\lambda^{jk}} \right] \Phi & =  \delta_{\xi^i(t)}\left(\lambda^i{}_j \xi^j(t) \right) \Phi \,.
\end{align}
Let us finally note that after the gauge fixing is performed, the only non-zero component of the dependent boost gauge field is given by $\omega_0{}^a$:
\begin{equation}
\omega_0{}^a = - \partial_a \Phi \,.
\end{equation}
Upon gauge fixing, one thus finds
\begin{equation}\label{eq:eom}
v^\mu f_\mu = 0 \qquad \Longrightarrow \qquad \partial^a \partial_a \Phi = 0 \,.
\end{equation}
One  thus sees that $v^\mu f_\mu = 0$ reduces to the Poisson equation upon gauge fixing.

\subsection{Gauging the Schr\"odinger algebra for $z\neq2$} \label{sec:sec23}

One can gauge the Schr\"odinger algebra for generic values of $z\neq1,2$ along similar lines as done above for the $z=2$ case. There are some differences with respect to the $z=2$ case stemming from the absence of the special conformal $K$ transformation and the fact that $N$ no longer corresponds to a central charge. In this section, we will gauge the Schr\"odinger algebra for generic values of $z$ and discuss the gauge-fixing procedure to accelerated extended Galilei transformations, paying special attention to these differences.

\subsubsection{Gauge transformations and constraints}\label{subsubsec:gaugetrafos-constraints-zneq2}

We will use the same notation for the gauge fields, parameters and covariant curvatures as in table \ref{table:zis2schr} (where however we now assume that the last line corresponding to the special conformal $K$ transformation is absent). By examining the Schr\"odinger algebra for $z\neq1,2$, the following gauge transformation rules for the gauge fields are found:
\begin{align}
\delta \tau_\mu & = z \Lambda_D \tau_\mu \,, \nonumber \\
\delta e_\mu{}^a & = \lambda^a \tau_\mu + \lambda^{a}{}_b e_{\mu}{}^{b} + \Lambda_D e_\mu{}^a \,, \nonumber\\
\delta \omega_\mu{}^{ab} & = \partial_\mu \lambda^{ab} + 2 \lambda^{c[a} \omega_\mu{}^{b]}{}_c\,,\nonumber \\
\delta \omega_\mu{}^a & = \partial_\mu \lambda^a + \lambda^a{}_b \omega_\mu{}^b + \lambda^b \omega_{\mu b}{}^a + (z-1) \lambda^a b_\mu - (z-1) \Lambda_D \omega_\mu{}^a \,, \nonumber \\
\delta b_\mu & = \partial_\mu \Lambda_D \,,\nonumber  \\
\delta m_\mu & = \partial_\mu \sigma + \lambda^a e_{\mu a} + (z-2) \sigma b_\mu -(z-2) \Lambda_D m_\mu \,,\label{eq:gaugetrafom}
\end{align}
where we have not indicated the transformation rules under $H$ and $P_a$ transformations, as they will be traded for general coordinate transformations later on, by imposing suitable constraints. The transformations of the inverse vielbeine are simply
\begin{align}\label{eq:trafoinvvielbeinszneq2}
\delta v^\mu & =  \lambda^a e^\mu{}_a-z\Lambda_D v^\mu\,,\nonumber \\
\delta e^\mu{}_a & =  \lambda_a{}^b e^\mu{}_b-\Lambda_D e^\mu{}_a\,.
\end{align}
Curvatures that are covariant with respect to \eqref{eq:gaugetrafom} are then given by:
\begin{align}\label{eq:curvatureszneq2}
R_{\mu \nu}(H) & = 2 \partial_{[\mu} \tau_{\nu]} - 2 z b_{[\mu} \tau_{\nu]} \,, \nonumber \\
R_{\mu \nu}{}^a(P) & = 2 \partial_{[\mu} e_{\nu]}{}^a - 2 \omega_{[\mu}{}^a \tau_{\nu]} - 2 \omega_{[\mu}{}^{ab} e_{\nu]b} - 2 b_{[\mu} e_{\nu]}{}^a \,, \nonumber \\
R_{\mu \nu}{}^{ab}(J) & = 2 \partial_{[\mu} \omega_{\nu]}{}^{ab} - 2 \omega_{[\mu}{}^{c[a} \omega_{\nu]}{}^{b]}{}_{c} \,, \nonumber \\
R_{\mu \nu}{}^a(G) & = 2 \partial_{[\mu} \omega_{\nu]}{}^a - 2 \omega_{[\mu}{}^{ab} \omega_{\nu]b} - 2 (z-1) \omega_{[\mu}{}^a b_{\nu]} \,, \nonumber \\
R_{\mu \nu}(D) & = 2 \partial_{[\mu} b_{\nu]} \,, \nonumber \\
R_{\mu \nu}(N) & = 2 \partial_{[\mu} m_{\nu]} - 2 \omega_{[\mu}{}^a e_{\nu]a} + 2 (z-2) b_{[\mu} m_{\nu]} \,.
\end{align}

As before, suitable constraints have to be introduced, whose purpose is to identify general coordinate transformations with local time and space translations and to turn some of the gauge fields into dependent ones. In particular, we start by imposing the following constraints:
\begin{equation}\label{eq:constraintsgeneralz}
R_{\mu \nu}(H) = 0 \,, \qquad R_{\mu \nu}{}^a(P) = 0 \,, \qquad R_{\mu \nu}(N) = 0 \,.
\end{equation}
These constraints imply that $H$ and $P_a$ transformations can be written as a combination of general coordinate transformations and other local Schr\"odinger transformations. 
The constraints of \eqref{eq:constraintsgeneralz} can also be used to make $\omega_\mu{}^{ab}$, $\omega_\mu{}^a$ and (the spatial) part of $b_\mu$ dependent. The solutions are as follows:
\begin{align}
\omega_\mu{}^{ab} & = -\frac12 \tau_\mu e^{\nu a} e^{\rho b} \left( 2 \partial_{[\nu} m_{\rho]} + 2 (z-2) b_{[\nu} m_{\rho]} \right) + e_\mu{}^c e^{\nu a} e^{\rho b} \partial_{[\nu} e_{\rho]c} - e^{\nu a} \partial_{[\mu} e_{\nu]}{}^b \nonumber \\ & \quad \quad +  e^{\nu b} \partial_{[\mu} e_{\nu]}{}^a + 2 e_\mu{}^{[a} e^{|\nu| b]} b_\nu \,, \label{eq:omega-ab} \\
\omega_\mu{}^a & = -\frac12 \tau_\mu v^\nu e^{\rho a} \left( 2 \partial_{[\nu} m_{\rho]} + 2 (z-2) b_{[\nu} m_{\rho]} \right) + \frac12 e^{\nu a} \left(2 \partial_{[\mu} m_{\nu]} + 2 (z-2) b_{[\mu} m_{\nu]} \right) \nonumber \\ & \quad \quad + e_\mu{}^b v^\nu e^{\rho a} \partial_{[\nu} e_{\rho]b} - v^\nu \partial_{[\mu} e_{\nu]}{}^a - e_\mu{}^a \left( v^\nu b_\nu \right)\,, \label{eq:omega-a} \\
e^\mu{}_a b_\mu & = -\frac2z e^\mu{}_a v^\nu \partial_{[\mu} \tau_{\nu]} \,.\label{eq:b}
\end{align}
Bianchi identities can lead to extra constraints. Upon imposing \eqref{eq:constraintsgeneralz} in the Bianchi identities, one gets the following extra constraints:
\begin{alignat}{3}
R_{0[ab]}(G) &= 0 \,, & \qquad R_{[abc]}(G) &= 0 \,, & \qquad 2 R_{0[ab]}{}^c(J) - 2 R_{0[a}(D)\delta_{b]}^c &=R_{ab}{}^c(G) \,, \nonumber\\
R_{[abc]}{}^d(J) &= 0 \,, & \qquad R_{ab}(D) &= 0 \,.
\end{alignat}
Note that, differently from the $z=2$ case, we have not put the full dilatation curvature $R_{\mu \nu}(D)$ to zero, but only its spatial part. In the $z=2$ case, imposing $R_{\mu\nu}(D)=0$ yielded a conventional constraint, used to solve for $f_\mu$. Since there are no $K$ transformations for $z\neq1,2$ and hence no gauge field $f_\mu$, putting $R_{\mu \nu}(D)$ to zero completely is not necessary here.

At this point, one is left with independent fields $\tau_\mu$, $e_\mu{}^a$, $m_\mu$ and $v^\mu b_\mu$. In the $z=2$ case, the latter could be put to zero by gauge fixing the special conformal $K$ transformation. This is no longer possible for $z\neq1,2$ and $v^\mu b_\mu$ thus remains as an independent field. In section \ref{sec:Stueckelbergscalar} we will see that when we add a St\"uckelberg scalar for the $N$ transformations there appears an extra special conformal type symmetry after imposing the curvature constraints that allows us to remove $v^\mu b_\mu$ by gauge fixing this special conformal transformation.

\subsubsection{Gauge fixing to acceleration extended Galilei symmetries}

As in the $z=2$ case, one can perform a partial gauge fixing to a formulation in which the left-over transformations constitute the acceleration extended Galilei symmetries. We will again perform this gauge fixing procedure for the case in which the spatial sections of the Newton--Cartan space-time are flat, i.e. taking:
\begin{equation}
\omega_{\mu}{}^{ab} = 0\,,
\end{equation}
and thus fixing spatial rotations $\lambda^{ab}(x^\mu)$ to be constant. By taking $\tau_\mu = \delta_\mu^0$, we can again completely fix dilatations and fix $\xi^0(x^\mu)$ transformations to be constant. Upon taking this gauge-fixing condition, one finds from \eqref{eq:b} that the only non-zero component of $b_\mu$ is the temporal one:
\begin{equation}
b_\mu = - \delta_\mu^0 (v^\nu b_\nu)\,.
\end{equation} 
In the $z=2$ case, this component could be put to zero, fixing the special conformal transformation. For $z\neq2$, this is no longer possible and this component remains. 

As before, the transformations $\xi^a(x^\mu)$ can be partially fixed to transformations of the form \eqref{fixedxia}
by taking $e_i{}^a = \delta_i^a$  and the boosts can be fixed by putting $e_0{}^a = 0$, at the expense of introducing a compensating transformation \eqref{fixedlambda}. The condition $\omega_0{}^{ab} = 0$ still implies that $m_i$ can be written as the $\partial_i$-derivative of a field $m$, that transforms under the remaining transformations as in \eqref{deltam}. One can therefore impose a gauge fixing condition of the form \eqref{fixm} to partially fix the transformations with parameter $\sigma(x^\mu)$. We will for simplicity take
\begin{equation}
m=0 \,.
\end{equation}
This condition is then preserved under $\sigma(x^\mu)$ transformations of the form:
\begin{equation}
\sigma(x^\mu) = \sigma(t) + \partial_t \xi^i(t) x^i \,.
\end{equation}
Renaming $m_0$ as $\Phi$, one is left over with two fields $\Phi$ and $b_0$, that transform under the remaining acceleration extended Galilei symmetries as:
\begin{align}
\delta \Phi & = \xi^0 \partial_t \Phi + \xi^i(t) \partial_i \Phi + \partial_t^2 \xi^i(t) x^i + (z-2) \partial_t \xi^i(t) x^i b_0 - \lambda^i{}_j x^j \partial_i \Phi \nonumber \\ & \quad + \partial_t \sigma(t) + (z-2) \sigma(t) b_0 \,, \nonumber \\
\delta b_0 & = \xi^0 \partial_t b_0 + \xi^i(t) \partial_i b_0 - \lambda^i{}_j x^j \partial_i b_0 \,.
\end{align}
One can check that these transformations close the algebra of acceleration extended Galilei symmetries, given by \eqref{boscomm}, on $\Phi$ and $b_0$.

\section{Promoting the central charge to a St\"uckelberg symmetry} \label{sec:sec3}

In the previous section, we showed how gauging the Schr\"odinger algebra led to TTNC geometry. The geometry defined by this gauging is however different from how TTNC geometry appears in Lifshitz holography (with a bulk massive vector field). One crucial difference is the fact that in Lifshitz holography the central charge is promoted to a St\"uckelberg symmetry, namely, it is accompanied by a St\"uckelberg scalar that shifts under $N$. This St\"uckelberg field was absent in the discussion of the previous section. Other (not unrelated) differences between the previous section and the appearance of torsional Newton--Cartan geometry in holography were also remarked. For instance, we found that it was not possible to obtain an affine connection that was invariant under both $G_a$ and $N$ transformations. Moreover, for $z\neq2$ the gauging led to an extra component, not present in the holographic description of torsional Newton--Cartan geometry.

In this section, we will show how a vielbein formulation of TTNC geometry can be defined in the presence of a scalar that promotes the central charge to a St\"uckelberg symmetry. Since a scalar is not a gauge field, this formulation will not be obtained via a gauging procedure. Instead, our starting point will be inspired by the end result of the previous section, namely the formulation of TTNC geometry that is covariant with respect to general coordinate transformations and an internal algebra of rotations, boosts, dilatations, central charge and special conformal transformations. Dependent gauge connections for rotations, boosts, dilatations and special conformal transformations will be defined in a similar way as in the previous section, namely via the definition of field strengths that are covariant with respect to the internal algebra, along with constraints on these curvatures. The imposition of a fully covariant vielbein postulate will then allow us to define an affine connection that is invariant under both $G_a$ and $N$ transformations.  Importantly, we will note that, even for $z\neq2$, one can consistently include a special conformal transformation in the internal algebra. Doing so will allow us to gauge fix the superfluous component of section \ref{sec:sec23}. In this way, we will be able to fully reproduce the TTNC geometry of \cite{Christensen:2013lma,Christensen:2013rfa}. 

We will first discuss the $z=2$ case in \ref{sec:Stueckelbergscalar}, while the $z\neq2$ case will be discussed in  \ref{sec:Stueckelbergscalarzneq2}.

\subsection{The case $z=2$}\label{sec:Stueckelbergscalar}

In order to make contact with the description of TTNC geometry of \cite{Christensen:2013lma,Christensen:2013rfa}, we will start with the set of independent fields $\tau_\mu$, $e_\mu{}^a$ and $m_\mu$ of section \ref{sec:sec22}, that transform under the internal symmetries as
\begin{align} \label{set1}
\delta\tau_\mu & =  2\Lambda_D\tau_\mu\,,\nonumber \\
\delta e_\mu{}^a & =  \lambda^a{}_b e_\mu{}^b+\lambda^a\tau_\mu+\Lambda_De_\mu{}^a\,,\nonumber \\
\delta m_\mu & = \partial_\mu \sigma + \lambda^a e_{\mu a} \,,
\end{align}
along with a scalar field $\chi$, whose transformation under internal symmetries is given by:
\begin{equation} \label{trafochizis2}
\delta\chi= \sigma\,.
\end{equation}
Under diffeomorphisms, the fields in \eqref{set1} transform as covectors, whereas the field $\chi$ transforms as an ordinary scalar. The transformation rule of $\chi$ is thus such that it promotes the central charge $N$ to a St\"uckelberg symmetry. It is easy to see that the above transformation rules close a commutator algebra on all independent fields, including $\chi$.

We stress that adding the scalar $\chi$ does not amount to a rewriting of the gauge structure of the fields that sit inside the Schr\"odinger gauge connection $\mathcal{A}_\mu$ defined in \eqref{eq:Schconnection}, as it introduces an extra component. Indeed, one could now fix the central charge transformation by choosing the gauge $\chi = 0$, leaving one with a vector $m_\mu$ that no longer transforms under the central charge. Alternatively, one can observe that $m_\mu$ and $\partial_\mu\chi$ can be combined in the $N$ inert combination $M_\mu$ defined by
\begin{equation}\label{eq:defMz=2}
M_\mu=m_\mu-\partial_\mu\chi = -\cal{D}_\mu \chi \,,
\end{equation} 
where in the last equality, we have used the covariant derivative $\cal{D}_\mu$ that acting on $\chi$ is covariant with respect to the transformations \eqref{trafochizis2} and \eqref{set1}.
As we will demonstrate shortly the effect of adding $\chi$ will effectively amount to replacing $m_\mu$ everywhere by $M_\mu$ thus adding one extra component to the formalism. 

In order to make the internal symmetries manifest, it is useful to introduce gauge connections for the spatial rotations, boosts, dilatations and special conformal transformations. To avoid introducing extra independent fields, these extra gauge connections have to be dependent (or pure gauge) and as in the previous section, a useful way to define them is via the introduction of covariant curvatures and constraints on them. Here we will show how this can be done in such a way that the resulting expressions for the dependent gauge fields contain the vector $M_\mu$ (and not $m_\mu$) and thus manifestly do not transform under $N$.

In order to do this, we will start from the covariant curvatures of section \ref{sec:sec22}. The spatial components of the dependent field $b_\mu$ are as before obtained as the solutions of the constraint $R_{\mu \nu}(H) = 0$. The temporal component is not determined; since it can be set to zero by fixing $K$, it is however not an independent component but rather corresponds to a gauge degree of freedom. To properly define $\omega_\mu{}^a$ and $\omega_\mu{}^{ab}$, we note that the curvature $R_{\mu \nu}(N)$ of eq. \eqref{curvatureszis2} can be rewritten entirely in terms of $M_\mu$ as:
\begin{equation}
R_{\mu \nu}(N) = \partial_{[\mu}M_{\nu]}-\omega_{[\mu}{}^ae_{\nu]a} \,.
\end{equation}
The connections $\omega_\mu{}^a$ and $\omega_\mu{}^{ab}$ can then be defined as the solutions of the constraints $R_{\mu \nu}{}^a(P) = R_{\mu \nu}(N) = 0$ and are given by eqs. \eqref{eq:expressionsomegaconnections}, with $m_\mu$ replaced by $M_\mu$.

To define $f_\mu$, we will use the constraint $R_{\mu \nu}(D) = 0$, that defines the spatial part of $f_\mu$ as before. To define the temporal part of $f_\mu$, we can now use the following modification of  \eqref{eq:scalarconstraint}:
\begin{equation}\label{eq:scalarconstraint2}
R_{0a}{}^a(G)+2M^bR_{0a}{}^a{}_b(J)+M^bM^cR_{ba}{}^a{}_c(J) = 0\,,
\end{equation}
where $M^b=e^{\mu b}M_\mu$. This leads to 
\begin{eqnarray}\label{eq:vf2}
v^\mu f_\mu & = & \frac{1}{d}v^\mu e^\nu{}_a\left(\partial_\mu\omega_\nu{}^a-\partial_\nu\omega_\mu{}^a+\omega_\mu{}^b\omega_\nu{}^a{}_b-\omega_\nu{}^b\omega_\mu{}^a{}_b+b_\mu\omega_\nu{}^a-b_\nu\omega_\mu{}^a\right)\nonumber\\
&&+\frac{2}{d}v^\mu M^c R_{\mu a}{}^a{}_c(J)-\frac{1}{d}M^bM^cR_{ba}{}^a{}_c(J)\,.
\end{eqnarray}
It can then be checked that the transformation rules of $\omega_\mu{}^{ab}$, $\omega_\mu{}^a$, $b_\mu$, $f_\mu$ are as in \eqref{gaugetrafoszis2}. These dependent fields can thus indeed be used as gauge connections for internal spatial rotations, boosts, dilatations and special conformal transformations.

The affine connection can be discussed in a similar way. In particular, since $\tilde\Gamma^\rho_{\mu\nu}$ in equation \eqref{eq:tildeGamma} only depends on the curl of $m_\mu$ we can now also write
\begin{eqnarray}
\tilde\Gamma^\rho_{\mu \nu} &=& -\left(v^\rho-h^{\rho\sigma}M_\sigma\right)\left(\partial_\mu-2b_\mu\right)\tau_\nu+\frac{1}{2}h^{\rho\sigma}\left[\left(\partial_\mu-2b_\mu\right)\left(h_{\nu\sigma}-\tau_\nu M_\sigma-\tau_\sigma M_\nu\right)\right.\nonumber\\
&&\left.+\left(\partial_\nu-2b_\nu\right)\left(h_{\mu\sigma}-\tau_\mu M_\sigma-\tau_\sigma M_\mu\right)-\left(\partial_\sigma-2b_\sigma\right)\left(h_{\mu\nu}-\tau_\mu M_\nu-\tau_\nu M_\mu\right)\right]\,.
\end{eqnarray}
The TTNC affine connection $\Gamma^\rho_{\mu\nu}$ defined by `throwing away the $b_\mu$ terms' is then given by
\begin{eqnarray}
\Gamma^\rho_{\mu \nu} &=& -\left(v^\rho-h^{\rho\sigma}M_\sigma\right)\partial_\mu\tau_\nu+\frac{1}{2}h^{\rho\sigma}\left[\partial_\mu\left(h_{\nu\sigma}-\tau_\nu M_\sigma-\tau_\sigma M_\nu\right)\right.\nonumber\\
&&\left.+\partial_\nu\left(h_{\mu\sigma}-\tau_\mu M_\sigma-\tau_\sigma M_\mu\right)-\partial_\sigma\left(h_{\mu\nu}-\tau_\mu M_\nu-\tau_\nu M_\mu\right)\right]\,,\label{eq:Gamma2}
\end{eqnarray}
and is manifestly $J_{ab}$, $G_a$, $K$ and $N$ invariant. 
It is again metric compatible in that it satisfies \eqref{eq:metriccompatible1} and \eqref{eq:metriccompatible2}. For the case of vanishing torsion $\partial_\mu\tau_\nu-\partial_\nu\tau_\mu=0$ the connection \eqref{eq:Gamma2} agrees with the one of NC geometry \cite{Andringa:2010it} because the $\chi$ field drops out.

\subsubsection{The Newton potential}\label{subsubsec:NP}

It is interesting to repeat the gauge fixing procedure of section \ref{subsubsec:gaugefixingz=2} in the presence of the field $\chi$. This proceeds analogously and we again end up with  equations \eqref{eq:trafosm0andm} together with the following transformation rule for $\chi$
\begin{equation}\label{eq:gaugefixingchi}
\delta\chi=\xi^0\partial_t\chi+\xi^i(t)\partial_i\chi-\lambda^i{}_j x^j\partial_i\chi+\sigma(x^\mu)\,.
\end{equation}
One can define the following $\sigma(x^\mu)$ transformation invariants 
\begin{eqnarray}
M & = & m-\chi\,,\\
\Phi & = & m_0-\partial_t m=M_0-\partial_t M\,,\\
M_i & = & \partial_i M\,,
\end{eqnarray}
where $\Phi$ is the Newton potential as defined in \eqref{eq:deff}. 

The transformations of $M$ and $M_0$ are obtained from \eqref{eq:trafosm0andm} and \eqref{eq:gaugefixingchi} and read
\begin{eqnarray}
\delta M & = & \xi^0 \partial_t M+ \xi^i(t) \partial_i M - \lambda^i{}_j x^j \partial_i M - \partial_t \xi^i(t) x^i  + Y(t) \,, \label{deltaM}\nonumber \\
\delta M_0 & = & \xi^0 \partial_t M_0 + \xi^i(t) \partial_i M_0- \lambda^i{}_j x^j \partial_i M_0 + \partial_t \xi^i(t) \partial_i M \,.
\end{eqnarray}
Using that the choices \eqref{fixtau}, \eqref{eq:choicee} and \eqref{eq:fixboosts} imply that
\begin{equation}
v^\mu=-\delta^\mu_t\,,\qquad e^\mu_a=\delta^\mu_a\,,
\end{equation}
the Newton potential $\Phi$ can also suggestively be written as
\begin{equation}\label{eq:NP2}
\Phi=-v^\mu M_\mu+\frac{1}{2}h^{\mu\nu}M_\mu M_\nu-\left(-v^\mu \partial_\mu M+\frac{1}{2}h^{\mu\nu}\partial_\mu M \partial_\nu M\right)\,.
\end{equation}
The first part of \eqref{eq:NP2} will be denoted by $\tilde\Phi$,
\begin{equation}\label{eq:NP3}
\tilde\Phi=-v^\mu M_\mu+\frac{1}{2}h^{\mu\nu}M_\mu M_\nu\,,
\end{equation}
and transforms as a scalar. The second part of \eqref{eq:NP2}, i.e. the term $-v^\mu \partial_\mu M+\frac{1}{2}h^{\mu\nu}\partial_\mu M \partial_\nu M=\partial_t M+\frac{1}{2}\partial_i M\partial^i M$ transforms as
\begin{equation}
\delta\left(\partial_t M+\frac{1}{2}\partial_i M\partial^i M\right)=\mathcal{L}_\xi\left(\partial_t M+\frac{1}{2}\partial_i M\partial^i M\right)-\partial_t^2\xi^i(t) x^i+Y'(t)\,,
\end{equation}
where the Lie derivative is along $\xi^\mu=(\xi^0\,,\xi^i(t)-\lambda^i{}_jx^j)$. Hence it is the second term in parenthesis in \eqref{eq:NP2} that is responsible for the $-\partial_t^2\xi^i(t) x^i+Y'(t)$ part of the transformation of the Newton potential in \eqref{eq:trafoNP}. On the flat NC background of section \ref{subsubsec:gaugefixingz=2} the relation between $\Phi$ and $\tilde\Phi$ is 
\begin{equation}
\tilde\Phi=\Phi+\partial_t M+\frac{1}{2}\partial_i M\partial^i M\,.
\end{equation}

It is not obvious how to extend the notion of a Newton potential in the sense of $\Phi$ to an arbitrary curved background. It is however straightforward to use $\tilde\Phi$ as defined in \eqref{eq:NP3}. This is why with a slight abuse of terminology $\tilde\Phi$ in \cite{Hartong:2014pma,Hartong:2014oma} is referred to as the Newton potential.

\subsection{The case $z\neq 2$}\label{sec:Stueckelbergscalarzneq2}

\subsubsection{Promoting the central charge to a St\"uckelberg symmetry}

For general $z\neq2$, we will again start from independent fields $\tau_\mu$, $e_\mu{}^a$, $m_\mu$ that transform under internal spatial rotations, boosts, dilatations and central charge transformations as
\begin{align} \label{set11}
\delta \tau_\mu & = z \Lambda_D \tau_\mu \,, \nonumber \\
\delta e_\mu{}^a & = \lambda^a \tau_\mu + \lambda^{a}{}_b e_{\mu}{}^{b} + \Lambda_D e_\mu{}^a \,, \nonumber\\
\delta m_\mu & = \partial_\mu \sigma + \lambda^a e_{\mu a} + (z-2) \sigma b_\mu -(z-2) \Lambda_D m_\mu \,,\end{align}
along with a scalar field $\chi$ that transforms as
\begin{equation}\label{eq:trafochizneq2}
\delta\chi=\sigma-(z-2)\Lambda_D\chi\,.
\end{equation}
As before, $\tau_\mu$, $e_\mu{}^a$ and $m_\mu$ transform as one-forms under diffeomorphisms, while $\chi$ is an ordinary scalar field. As in the $z=2$ case, the addition of $\chi$ promotes $N$ to a St\"uckelberg symmetry and we can define a field denoted by $M_\mu$ that is inert under $N$ as follows\footnote{In the holographic context the field $\tilde m_\mu=m_\mu+(2-z)\chi b_\mu$ plays an important role as the source for the mass current \cite{Hartong:2014pma,Hartong:2014oma}.}
\begin{equation}\label{eq:defM}
M_\mu=m_\mu+(2-z)\chi b_\mu-\partial_\mu\chi = - \cal{D}_\mu \chi \,,
\end{equation}
where in the last equality, we have used the covariant derivative $\cal{D}_\mu$ that acting on $\chi$ is covariant with respect to the transformations \eqref{eq:trafochizneq2} and \eqref{set11}.
The field $M_\mu$ transforms as 
\begin{equation}\label{eq:trafoM}
\delta M_\mu=e_\mu{}^a\lambda_a+(2-z)\Lambda_D M_\mu\,.
\end{equation}
To be able to make the internal spatial rotations, boosts and dilatations manifest, we will introduce dependent gauge connections for them. To do this, we will again start from curvature constraints $R_{\mu \nu}(H)=0$, $R_{\mu\nu}{}^a(P)=0$, with $R_{\mu \nu}(H)$, $R_{\mu\nu}{}^a(P)$ defined in \eqref{eq:curvatureszneq2}, along with the constraint
\begin{equation}\label{eq:newcurvconstraint}
\partial_{[\mu} M_{\nu]}-(2-z)b_{[\mu} M_{\nu]}-\omega_{[\mu}{}^a e_{\nu] a}=0\,.
\end{equation}
Solving $R_{\mu \nu}{}^a(P) = 0$ and \eqref{eq:newcurvconstraint} leads to expressions for $\omega_\mu^{ab}$ and $\omega_\mu{}^a$, given by equations \eqref{eq:omega-ab}, \eqref{eq:omega-a} with $m_\mu$ replaced by $M_\mu$. The constraint $R_{\mu \nu}(H) = 0$ can be used to solve for the spatial part of $b_\mu$ (given by \eqref{eq:b}). Note that at this stage the temporal part of $b_\mu$ is not determined yet. In the next subsection, we will show that one can consistently extend the internal symmetries with a special conformal transformation, that allows one to gauge fix this component to zero, as in the $z=2$ case.

One can now impose vielbein postulates that are covariant with respect to the internal symmetries:
\begin{align} \label{vielbpostszneq2}
\mathcal{D}_\mu \tau_\nu &\equiv \partial_\mu \tau_\nu - \tilde\Gamma^\rho_{\mu \nu} \tau_\rho -zb_\mu\tau_\nu= 0 \,, \nonumber \\
\mathcal{D}_\mu e_\nu{}^a &\equiv \partial_\mu e_\nu{}^a - \tilde\Gamma^\rho_{\mu \nu} e_\rho{}^a - \omega_\mu{}^a{}_b e_\nu{}^b - \omega_\mu{}^a \tau_\nu-b_\mu e_\nu{}^a = 0 \,,
\end{align}
where $\omega_\mu{}^{ab}$ and $\omega_\mu{}^a$ are given by \eqref{eq:omega-ab} and \eqref{eq:omega-a} with $m_\mu$ replaced by $M_\mu$. Solving for $\tilde\Gamma^\rho_{\mu \nu}$ and using \eqref{eq:b} we can write the solution as
\begin{eqnarray}
\tilde\Gamma^\rho_{\mu \nu} &=& -\left(v^\rho-h^{\rho\sigma}M_\sigma\right)\left(\partial_\mu-zb_\mu\right)\tau_\nu+\frac{1}{2}h^{\rho\sigma}\left[\left(\partial_\mu-2b_\mu\right)\left(h_{\nu\sigma}-\tau_\nu M_\sigma-\tau_\sigma M_\nu\right)\right.\nonumber\\
&&\left.+\left(\partial_\nu-2b_\nu\right)\left(h_{\mu\sigma}-\tau_\mu M_\sigma-\tau_\sigma M_\mu\right)-\left(\partial_\sigma-2b_\sigma\right)\left(h_{\mu\nu}-\tau_\mu M_\nu-\tau_\nu M_\mu\right)\right]\,.\label{eq:tildeGammazneq2}
\end{eqnarray}
Dropping the $b_\mu$ terms thus leaves us with the $G_a$, $J_{ab}$ and $N$ invariant affine TTNC connection $\Gamma^\rho_{\mu\nu}$ given in \eqref{eq:Gamma2} (albeit with a different assignment of dilatation weights to the various fields appearing in $\Gamma^\rho_{\mu\nu}$). Again $\tilde\Gamma^\rho_{\mu \nu}$ is symmetric by virtue of the $R_{\mu\nu}(H)=0$ constraint. We can also define the covariant derivative of $M_\mu$ as
\begin{equation}
\mathcal{D}_\mu M_\nu=\partial_\mu M_\nu-\tilde\Gamma^{\rho}_{\mu\nu}M_\rho-(2-z)b_\mu M_\nu-\omega_\mu{}^a e_{\nu a}\,.
\end{equation}
The curvature constraints $R_{\mu\nu}(H)=0$, $R_{\mu\nu}{}^a(P)=0$ and \eqref{eq:newcurvconstraint} can then all be rewritten as:
\begin{equation}\label{eq:curvatureconstraintsv2.0}
\mathcal{D}_{[\mu}\tau_{\nu]}=0\,,\qquad\mathcal{D}_{[\mu}e_{\nu]}{}^a=0\,,\qquad\mathcal{D}_{[\mu}M_{\nu]}=0\,.
\end{equation}

\subsubsection{Adding a special conformal symmetry}\label{subsec:enhancement}

The TTNC geometry is formulated in terms of the fields $\tau_\mu$, $e_\mu{}^a$ and $M_\mu$. In the previous subsection, we have introduced a gauge field $b_\mu$ for dilatations, of which the spatial part is dependent. The temporal part $v^\mu b_\mu$ is however still undetermined. As we do not wish to include it as an independent field in the formulation of TTNC geometry, we should either make it dependent or turn it into a gauge degree of freedom. Here we will argue that the latter is possible, i.e. like for $z=2$ one can add a special conformal symmetry $K$ that acts on $b_\mu$ as:
\begin{equation}\label{eq:deltaKonb}
\delta_K b_\mu=\Lambda_K\tau_\mu\,.
\end{equation}
The component $v^\mu b_\mu$ is then a gauge degree of freedom and can be set to zero by fixing this local $K$ transformation. As the fields describing the TTNC geometry are required to remain inert under this symmetry, we conclude from \eqref{eq:defM} that the independent field $m_\mu$ transforms under $K$ as
\begin{equation}\label{eq:deltaKonm}
\delta_K m_\mu=(z-2)\Lambda_K\chi\tau_\mu\,.
\end{equation}
Using the explicit expressions for the dependent connections introduced so far, i.e. using \eqref{eq:omega-ab} and \eqref{eq:omega-a} with $m_\mu$ replaced by $M_\mu$, we find that  $\omega_\mu{}^a$ also transforms under this $K$ transformation, according to
\begin{equation}\label{eq:deltaKomegaboost}
\delta_K\omega_\mu{}^a=(z-1)\Lambda_Ke_\mu{}^a-(z-2)\Lambda_K\left(e_\mu{}^a-\tau_\mu e^{\rho a}M_\rho\right)=\Lambda_K\left(e_\mu{}^a+(z-2)\tau_\mu M^a\right)\,.
\end{equation}
Let us stress that for $z\neq 2$ this symmetry is not part of the Schr\"odinger algebra. Since here, we are not considering the full Schr\"odinger algebra, but only the internal part (i.e. not including space-time translations), it is however possible to add this special conformal symmetry by hand in a consistent manner. Acting on $m_\mu$, this transformation is not of the Yang--Mills form since it includes a non-linear term $\chi \tau_\mu$, involving the scalar $\chi$. Note however that for $z=2$ the above $K$ transformations agree with the way the special conformal generator $K$ acts on the various fields introduced so far.

We have thus added an extra $K$ symmetry. In order to make its presence manifest, we can again introduce a  gauge connection $f_\mu$ for this symmetry, that should be dependent in order not to introduce new independent components to the formalism. The precise expression will be derived in the next section. Let us here however determine the transformation law that we wish this dependent gauge field to obey. In this respect, we note that the addition of the extra $K$ transformation implies that the field strength of $b_\mu$, that was introduced as a covariant curvature in \eqref{eq:curvatureszneq2} no longer transforms covariantly. We can remedy this by defining a new $R_{\mu \nu}(D)$ curvature, that includes additional $f_\mu$ terms for $K$ transformations:
\begin{equation}\label{eq:newDcurv}
R_{\mu\nu}(D)=\partial_\mu b_\nu-\partial_\nu b_\mu-f_\mu\tau_\nu+f_\nu\tau_\mu\,.
\end{equation}
Requiring that this new curvature transforms covariantly as
\begin{equation}\label{eq:trafonewDcurv}
\delta R_{\mu\nu}(D)=\Lambda_K\left(\partial_\mu\tau_\nu-\partial_\nu\tau_\mu-zb_\mu\tau_\nu+zb_\nu\tau_\mu\right)= 0\,,
\end{equation}
by virtue of the first constraint in \eqref{eq:curvatureconstraintsv2.0}, implies that $f_\mu$ must transform under internal symmetries as
\begin{equation}\label{eq:trafo-f}
\delta f_\mu=\partial_\mu\Lambda_K-z\Lambda_D f_\mu+z\Lambda_K b_\mu\,,
\end{equation}
where we have used that
\begin{equation}
\delta b_\mu=\partial_\mu\Lambda_D+\Lambda_K\tau_\mu\,.
\end{equation}
Other curvatures that appeared in \eqref{eq:curvatureszneq2} also cease to transform covariantly under internal symmetries, once $K$ has been introduced. Correct covariant curvatures can be defined 
by considering the commutators on $\chi$ and $M^a$ given by
\begin{eqnarray}
\left[\mathcal{D}_\mu\,,\mathcal{D}_\nu\right]\chi & = & -R_{\mu\nu}(N)-(2-z)\chi  R_{\mu\nu}(D)\,,\\
\left[\mathcal{D}_\mu\,,\mathcal{D}_\nu\right] M^a & = & -R_{\mu\nu}{}^{ab}(J)M_b-(1-z) R_{\mu\nu}(D)M^a- R_{\mu\nu}{}^a(G)\,,\label{eq:commutatorMa}
\end{eqnarray}
where $\mathcal{D}_\mu$ is covariant with respect to both diffeomorphisms (that drop out of commutators) and $D$, $G_a$, $J_{ab}$, $N$, $K$ transformations. The curvatures thus defined are given by
\begin{eqnarray}
 R_{\mu\nu}(D) & = & \partial_\mu b_\nu-\partial_\nu b_\mu-f_\mu\tau_\nu+f_\nu\tau_\mu\,,\\
 R_{\mu\nu}(N) & = & 2\partial_{[\mu}m_{\nu]}-2(2-z)b_{[\mu}m_{\nu]}-2\omega_{[\mu}{}^a e_{\nu]a}+2(2-z)\chi f_{[\mu}\tau_{\nu]}\,,\\
R_{\mu\nu}{}^{ab}(J) & = & 2\partial_{[\mu} \omega_{\nu]}{}^{ab}-2\omega_{[\mu}{}^{ca}\omega_{\nu]}{}^b{}_c\,,\\
 R_{\mu\nu}{}^a(G) & = & 2\partial_{[\mu}\omega_{\nu]}{}^a-2\omega_{[\mu}{}^{ab}\omega_{\nu]b}-2(1-z)b_{[\mu}\omega_{\nu]}{}^a\nonumber\\
&&-2f_{[\mu}\left(e_{\nu]}{}^a+(z-2)\tau_{\nu]}M^a\right)\,.\label{eq:newGcurv}
\end{eqnarray}
Finally one can also define a curvature for $f_\mu$ via
\begin{equation}
R_{\mu\nu}(K)=\partial_\mu f_\nu-\partial_\nu f_\mu+zb_\mu f_\nu -z b_\nu f_\mu\,.
\end{equation}
Using some of these curvature definitions, we will now derive an expression for the dependent gauge field $f_\mu$ that transforms as in \eqref{eq:trafo-f}.

\subsubsection{The dependent gauge connection $f_\mu$}

In order to construct a fully dependent gauge connection $f_\mu$ that transforms as in \eqref{eq:trafo-f}, we start by imposing the curvature constraint $R_{\mu\nu}(D)=0$ from which we find
\begin{equation}
f_\mu=v^\rho\left(\partial_\rho b_\mu-\partial_\mu b_\rho\right)-v^\rho f_\rho\tau_\mu\,.
\end{equation}
We thus need to find an expression for $v^\rho f_\rho$ which transform as
\begin{equation}
\delta\left(v^\mu f_\mu\right)=\lambda^\mu f_\mu+v^\mu\partial_\mu\Lambda_K-2z\Lambda_D v^\mu f_\mu+z\Lambda_K v^\mu b_\mu\,,
\end{equation}
in order that $f_\mu$ transforms as \eqref{eq:trafo-f}.

Before embarking on the construction of $v^\mu f_\mu$ we first derive a Bianchi identity that will prove useful later. Multiplying \eqref{eq:commutatorMa} by $-\tau_\rho$, antisymmetrizing over all indices and using the vielbein postulates \eqref{vielbpostszneq2} together with the identity
\begin{equation}
2\mathcal{D}_{[\mu}\mathcal{D}_{\nu}\left(e_{\rho]}{}^a-\tau_{\rho]}M^a\right)=-R_{[\mu\nu}{}^{ab}(J)\left(e_{\rho]b}-\tau_{\rho]}M^b\right)- R_{[\mu\nu}(D)\left(e_{\rho]}{}^a-\tau_{\rho]}M^a\right)\,,
\end{equation}
where we used the fact that $e_{\mu}{}^a-\tau_{\mu}M^a$ is boost invariant,
we obtain the Bianchi identity
\begin{equation}\label{eq:BI}
R_{[\mu\nu}{}^{ab}(J)e_{\rho]b}+ R_{[\mu\nu}{}^a(G)\tau_{\rho]}=0\,.
\end{equation}
By contracting this with $v^\mu e^\nu{}_c e^\rho{}_a$ we find
\begin{equation}\label{eq:BII}
 R_{ca}{}^a(G) + v^\mu R_{\mu a}{}^a{}_c(J) = 0\,,
\end{equation}
and by contracting \eqref{eq:BI} with $e^\mu{}_b e^\nu{}_a e^\rho{}_c$ we obtain
\begin{equation}
R_{ba}{}^a{}_c(J)-R_{ca}{}^a{}_b(J) = 0\,.
\end{equation}

Inspired by the $z=2$ result we make the follow ansatz for $v^\mu f_\mu$
\begin{eqnarray}
v^\mu f_\mu & = & F+\frac{1}{d}v^\mu e^\nu{}_a\left(2\partial_{[\mu}\omega_{\nu]}{}^a-2\omega_{[\mu}{}^{ab}\omega_{\nu]b}-2(1-z)b_{[\mu}\omega_{\nu]}{}^a-2(z-2)f_{[\mu}\tau_{\nu]}M^a\right)\,,
\end{eqnarray}
where $F$ needs to be determined. Note that the right hand side does not contain $v^\mu f_\mu$ but only $e^\mu{}_a f_\mu$ which we already know. This transforms under boosts as
\begin{equation}
\delta_G \left(v^\mu f_\mu\right)=\delta_G F+\lambda^a e^\mu{}_a f_\mu+\frac{1}{d}\lambda^c\left[ R_{ca}{}^a(G)-v^\mu R_{\mu a}{}^a{}_c(J)\right]\,.
\end{equation} 
In order to cancel the curvature terms upon use of the Bianchi identities \eqref{eq:BI} and \eqref{eq:BII} we take for $F$
\begin{equation}
F=\frac{2}{d}v^\mu M^c R_{\mu a}{}^a{}_c(J)-\frac{1}{d}M^b M^c R_{ba}{}^a{}_c(J)+\tilde F\,.
\end{equation}
This leads to 
\begin{equation}
\delta_G \left(v^\mu f_\mu\right)=\delta_G\tilde F+\lambda^a e^\mu{}_a f_\mu\,.
\end{equation}
We will take $\delta_G\tilde F=0$ so that we obtain a boost invariant expression for $f_\mu$. We still need to ensure that $f_\mu$ transforms as in \eqref{eq:trafo-f} with respect to dilatations and special conformal transformations. It is straightforward to check that under dilatations we have
\begin{equation}
\delta_D\left(v^\mu f_\mu\right)=-2z\Lambda_D v^\mu f_\mu
\end{equation}
provided we take $\delta_D\tilde F=-2z\Lambda_D\tilde F$. Finally under special conformal transformations we have
\begin{equation}
\delta_K\left(v^\mu f_\mu\right)=\delta_K\tilde F+v^\mu\partial_\mu\Lambda_K+z\Lambda_K v^\mu b_\mu+\frac{1}{d}(z-2)\Lambda_K e^\mu{}_a\mathcal{D}_\mu M^a\,.
\end{equation}
The last term can be cancelled by taking
\begin{equation}
\tilde F=\frac{1}{2d^2}(z-2)\left(e^\mu{}_a\mathcal{D}_\mu M^a\right)^2\,,
\end{equation}
where 
\begin{equation}
e^\mu{}_a\mathcal{D}_\mu M^a=-\frac{1}{e}\partial_\mu\left[e\left(v^\mu-e^\mu_aM^a\right)\right]+d\left(v^\mu-e^\mu_aM^a\right) b_\mu\,,
\end{equation}
with $e$ the determinant of the matrix formed by $(\tau_\mu\,, e_\mu{}^a)$. One sees that $e^\mu{}_a\mathcal{D}_\mu M^a$ is boost invariant and has dilatation weight $-z$ so that $\tilde F$ obeys the conditions $\delta_G\tilde F=0$ and $\delta_D\tilde F=-2z\Lambda_D\tilde F$. The final expression for $v^\mu f_\mu$ is thus
\begin{eqnarray}
v^\mu f_\mu & = & \frac{1}{d}v^\mu e^\nu{}_a\left(2\partial_{[\mu}\omega_{\nu]}{}^a-2\omega_{[\mu}{}^{ab}\omega_{\nu]b}-2(1-z)b_{[\mu}\omega_{\nu]}{}^a-2(z-2)f_{[\mu}\tau_{\nu]}M^a\right)\nonumber\\
&&+\frac{2}{d}v^\mu M^c R_{\mu a}{}^a{}_c(J)-\frac{1}{d}M^b M^c R_{ba}{}^a{}_c(J)+\frac{1}{2d^2}(z-2)\left(e^\mu{}_a\mathcal{D}_\mu M^a\right)^2\,.
\end{eqnarray}
We thus see that $f_\mu$ is a completely dependent gauge connection.

\subsubsection{Summary of fields and transformation rules}

In summary, for $z\neq2$, we have the following set of fields and local transformations:
\begin{eqnarray}
\delta \tau_\mu & = &z \Lambda_D \tau_\mu \,,\label{eq:trafotau2} \\
\delta e_\mu{}^a & = & \tau_\mu\lambda^a + \lambda^{a}{}_b e_{\mu}{}^{b} + \Lambda_D e_\mu{}^a \,,\label{eq:trafoe2}\\
\delta m_\mu & = & \partial_\mu \sigma + \lambda^a e_{\mu a} - (2-z) \sigma b_\mu +(2-z) \Lambda_D m_\mu-(2-z)\Lambda_K\chi\tau_\mu\,,\\
\delta\omega_\mu{}^a & = & \partial_\mu\lambda^a+(z-1)\lambda^a b_\mu-\lambda^b\omega_\mu{}^a{}_b-(z-1)\Lambda_D\omega_\mu{}^a+\lambda^a{}_b\omega_\mu{}^b\nonumber\\
&&+\Lambda_K\left(e_\mu{}^a+(z-2)\tau_\mu M^a\right)\,,\\
\delta\omega_\mu{}^{ab} & = & \partial_\mu\lambda^{ab}+2\lambda^{c[a}\omega_{\mu}{}^{b]}{}_c\,,\\
\delta b_\mu & = & \partial_\mu\Lambda_D+\Lambda_K\tau_\mu\,,\\
\delta f_\mu & = & \partial_\mu\Lambda_K-z\Lambda_D f_\mu+z\Lambda_K b_\mu\,,\\
\delta \chi & = & \sigma+(2-z)\Lambda_D\chi\,,\label{eq:trafochi2}
\end{eqnarray}
where we have only written the transformations for internal symmetries (boosts, rotations, central charge gauge transformations, dilatations and special conformal transformations) and have left out the diffeomorphisms. Only the fields $\tau_\mu$, $e_\mu{}^a$ and $M_\mu = m_\mu - \partial_\mu \chi$ are independent. The other fields are dependent or pure gauge and in case they are dependent, they can be defined via appropriate constraints on curvatures that are covariant with respect to the above transformations. The last terms in $\delta m_\mu$ and $\delta\omega_\mu{}^a$ describe the coupling to $\chi$ under special conformal transformations. One can check that this set of local transformations forms a closed algebra. For $z=2$ the transformations agree with those of the $D$, $G_a$, $J_{ab}$, $N$, $K$ subgroup of the $z=2$ Schr\"odinger group. 

In this way, we have obtained a description of TTNC geometry for $z\neq2$, that makes the presence of Schr\"odinger-type symmetries manifest. It is given in terms of independent fields $\tau_\mu$, $e_\mu{}^a$ and $M_\mu$ with dilatation weights $z$, $1$ and $2-z$, respectively, where the field $\tau_\mu$ is hypersurface orthogonal. In the next section we will generalize these results to the case of TNC geometry where there are no constraints imposed on $\tau_\mu$.

\section{Torsional Newton--Cartan geometry}\label{sec:TNC}

In \cite{Hartong:2014pma,Hartong:2014oma} it is shown that asymptotically locally Lifshitz space-times with dynamical critical exponent $z$ in the range $1<z\le 2$ are dual to field theories that live on a torsional Newton--Cartan space-time where typically $\tau_\mu$ is fully unconstrained. Only for certain special $z=2$ cases (and whenever $z>2$) one finds that $\tau_\mu$ is hypersurface orthogonal so that the boundary geometry becomes TTNC. Here we will describe the geometry one obtains if we generalize TTNC to the case where $\tau_\mu$ is fully unconstrained. The resulting geometry is called torsional Newton--Cartan (TNC) geometry \cite{Christensen:2013lma,Christensen:2013rfa}.

As shown in \cite{Hartong:2014pma,Hartong:2014oma} the holographic boundary data are provided by the fields: $\tau_\mu$, $e_\mu{}^a$, $M_\mu$ and $\chi$ transforming as in \eqref{eq:gaugetrafom}, \eqref{eq:trafochizneq2} and \eqref{eq:trafoM} with generally no constraint on $\tau_\mu$. Below we will work out the properties of the TNC geometry and show that the result can be viewed as adding torsion to the results of the previous section, in the sense that the connection $\tilde\Gamma^\rho_{\mu\nu}$ which is torsionless in the case of TTNC geometry now acquires torsion.

\subsection{Invariants}

The first step in setting up the TNC geometry is the construction of invariants. By this we mean tensors with a specific dilatation weight that are invariant under $G_a$, $J_{ab}$ and $N$ transformations. These invariants are given by
\begin{eqnarray}
\hat v^\mu & = & v^\mu-h^{\mu\nu}M_\nu\,,\\
\bar h_{\mu\nu} & = & h_{\mu\nu}-\tau_\mu M_\nu-\tau_\nu M_\mu\,,\\
\tilde\Phi & = & -v^\mu M_\mu+\frac{1}{2}h^{\mu\nu}M_\mu M_\nu\,,\label{eq:tildePhi}
\end{eqnarray}
together with the degenerate metric invariants $\tau_\mu$ and $h^{\mu\nu}$. The quantity $\bar h_{\mu\nu}$ appeared earlier in the construction of $\Gamma^\rho_{\mu\nu}$ (section \ref{subsec:affineconnection}) whereas $\tilde\Phi$ appeared already in section \ref{subsubsec:NP}. Their dilatation weights are given in table \ref{table:dimensionsinvariants}.
\begin{table}[h!]
      \centering
      \begin{tabular}{|c|c|c|c|c|c|}
      \hline
invariant & $\tau_\mu$ & $\bar h_{\mu\nu}$ & $\tilde\Phi$ & $\hat v^\mu$ & $h^{\mu\nu}$ \\
  \hline
 dilatation weight & $z$ & $2$ & $-2(z-1)$ & $-z$ & $-2$    \\
         \hline
           \end{tabular}
      \caption{Dilatation weights of the TNC invariants.}\label{table:dimensionsinvariants}
\end{table}

It will also sometimes be useful to use the $G_a$ and $N$ invariant vielbein $\hat e_\mu{}^a$ defined as
\begin{equation}\label{eq:hate}
\hat e_\mu{}^a = e_\mu{}^a-\tau_\mu M^a\,.
\end{equation}
The objects $\hat e_\mu{}^a$, $\hat v^\mu$, $\tau_\mu$ and $e_\mu{}^a$ form an orthonormal set, i.e.
\begin{alignat}{3}
\hat v^\mu \tau_\mu &= -1\,, & \qquad \qquad  \hat v^\mu \hat e_\mu{}^a &= 0 \,, & \qquad \qquad \tau_\mu e^\mu{}_a &= 0\,,\nonumber \\
\hat e_\mu{}^a e^\mu{}_b &= \delta^a_b\,, & e^\mu{}_a \hat e^a{}_\nu & = \delta^\mu_\nu + \hat v^\mu \tau_\nu \,.
\end{alignat}

\subsection{Vielbein postulates}

The $G_a$, $J_{ab}$ and $N$ invariant affine connection that is metric compatible in the sense that
\begin{eqnarray}
\nabla_\mu\tau_\nu & = & 0\,,\label{eq:TNC1}\\
\nabla_\mu h^{\nu\rho} & = & 0\,,\label{eq:TNC2}
\end{eqnarray}
is the same as found before in equation \eqref{eq:Gamma2} (although here there are no constraints imposed on $\tau_\mu$) which we repeat here for convenience
\begin{equation}\label{eq:GammaTNC}
\Gamma^{\rho}_{\mu\nu} = -\hat v^\rho\partial_\mu\tau_\nu+\frac{1}{2}h^{\rho\sigma}\left(\partial_\mu\bar h_{\nu\sigma}+\partial_\nu \bar h_{\mu\sigma}-\partial_\sigma\bar h_{\mu\nu}\right)\,,
\end{equation}
and is a simple expression in terms of the invariants. 

The approach that we take in deriving the properties of TNC geometry is reversed to the approach taken before when gauging the Schr\"odinger algebra. In the latter case we guessed the relevant group, gauged it and via a vielbein postulate found $\Gamma^\rho_{\mu\nu}$. Here we guess $\Gamma^\rho_{\mu\nu}$ and we work our way towards unraveling the underlying Schr\"odinger symmetries.

Define the following covariant derivatives
\begin{eqnarray}
\mathcal{D}_\mu\tau_\nu & = &\partial_\mu\tau_\nu-\Gamma^\rho_{\mu\nu}\tau_\rho\,,\label{eq:covder1}  \\
\mathcal{D}_\mu e_\nu{}^a & = & \partial_\mu e_\nu{}^a-\Gamma^\rho_{\mu\nu}e_\rho{}^a-\Omega_\mu{}^a\tau_\nu-\Omega_\mu{}^{a}{}_be_{\nu}{}^b \,, \\
\mathcal{D}_\mu v^\nu & = & \partial_\mu v^\nu+\Gamma^\nu_{\mu\rho}v^\rho-\Omega_\mu{}^a e^\nu{}_a\,,  \\
\mathcal{D}_\mu e^\nu{}_a & = & \partial_\mu e^\nu{}_a+\Gamma^\nu_{\mu\rho}e^\rho{}_a+\Omega_\mu{}^{b}{}_ae^\nu{}_b \,, \label{eq:covder4}
\end{eqnarray}
compatible with \eqref{eq:TNC1} and \eqref{eq:TNC2} and the transformations of $\tau_\mu$ and $e_\mu{}^a$ given in \eqref{eq:gaugetrafom} as well as those of the inverse vielbeine given in \eqref{eq:trafoinvvielbeinszneq2}. Impose the following vielbein postulates
\begin{eqnarray}
\mathcal{D}_\mu\tau_\nu & = & 0 \,, \label{eq:VP1} \\
\mathcal{D}_\mu e_\nu{}^a & = & 0 \,, \\
\mathcal{D}_\mu v^\nu & = & 0\,,\\
\mathcal{D}_\mu e^\nu{}_a & = & 0\,,\label{eq:VP4}
\end{eqnarray}
and take $\Gamma^\rho_{\mu\nu}$ as in \eqref{eq:GammaTNC}. The connections $\Omega_\mu{}^a$ and $\Omega_\mu{}^{ab}$ can be solved for in terms of $\Gamma^\rho_{\mu\nu}$ and they are the non-dilatation covariant boost and rotation connections, respectively. Their relation with $\omega_\mu{}^a$ and $\omega_\mu{}^{ab}$ for the case of TTNC geometry will become clear shortly.

\subsection{The dilatation connection $b_\mu$}

To make local dilatation covariance manifest we use a procedure that we call anisotropic Weyl-gauging which is a straightforward generalization to $z>1$ of the Weyl gauging technique used in relativistic settings. The main ingredient is a dilatation connection $b_{\mu}$ that transforms under dilatations as 
\begin{equation}
\delta_D b_\mu=\partial_\mu\Lambda_D\,,
\end{equation}
and that is invariant under the $G_a$, $J_{ab}$, $N$ transformations (without imposing any constraint on $\tau_\mu$). Inspired by the TTNC connection \eqref{eq:b} for dilatations we define here $b_\mu$ in terms of the invariants as follows
\begin{equation}\label{eq:TNCb}
b_\mu=\frac{1}{z}\hat v^\rho\left(\partial_\rho\tau_\mu-\partial_\mu\tau_\rho\right)-\hat v^\rho b_\rho \tau_\mu\,.
\end{equation}

We will use this $b_\mu$ field to rewrite the covariant derivatives \eqref{eq:covder1}--\eqref{eq:covder4} in a manifestly dilatation covariant manner. To do this we take $\Gamma^\rho_{\mu\nu}$ of equation \eqref{eq:GammaTNC} and replace ordinary derivatives by dilatation covariant ones leading to a new connection $\tilde\Gamma^{\rho}_{\mu\nu}$ that is invariant under the $G_a$, $J_{ab}$, $N$ and $D$ transformations and reads
\begin{equation}\label{eq:tildeGammaTNC}
\tilde\Gamma^{\rho}_{\mu\nu} = -\hat v^\rho\left(\partial_\mu-zb_\mu\right)\tau_\nu+\frac{1}{2}h^{\rho\sigma}\left(\left(\partial_\mu-2b_\mu\right)\bar h_{\nu\sigma}+\left(\partial_\nu-2b_\nu\right) \bar h_{\mu\sigma}-\left(\partial_\sigma-2b_\sigma\right)\bar h_{\mu\nu}\right)\,.
\end{equation}
This is the same expression as given earlier in the case of TTNC geometry, equation \eqref{eq:tildeGammazneq2} except that now of course $\tau_\mu$ is not constrained to be hypersurface orthogonal. With the help of $b_\mu$ and $\tilde\Gamma^{\rho}_{\mu\nu}$ we can now rewrite the covariant derivatives \eqref{eq:covder1}--\eqref{eq:covder4} as follows
\begin{eqnarray}
\mathcal{D}_\mu\tau_\nu & = &\partial_\mu\tau_\nu-\tilde\Gamma^{\rho}_{\mu\nu}\tau_\rho-zb_\mu\tau_\nu\,,\label{eq:covder1v2}  \\
\mathcal{D}_\mu e_\nu{}^a & = & \partial_\mu e_\nu{}^a-\tilde\Gamma^{\rho}_{\mu\nu}e_\rho{}^a-\omega_\mu{}^a\tau_\nu-\omega_\mu{}^{a}{}_be_{\nu}{}^b-b_\mu e_\nu{}^a \,, \label{eq:covder2v2} \\
\mathcal{D}_\mu v^\nu & = & \partial_\mu v^\nu+\tilde\Gamma^{\nu}_{\mu\rho}v^\rho-\omega_\mu{}^a e^\nu{}_a+zb_\mu v^\nu\,,  \\
\mathcal{D}_\mu e^\nu{}_a & = & \partial_\mu e^\nu{}_a+\tilde\Gamma^{\nu}_{\mu\rho}e^\rho{}_a+\omega_\mu{}^{b}{}_ae^\nu{}_b +b_\mu e^\nu{}_a\,,\label{eq:covder4v2}
\end{eqnarray}
where the $\omega_\mu{}^a$ and $\omega_\mu{}^{ab}$ connections are written in terms of $\Omega_\mu{}^a$ and $\Omega_\mu{}^{ab}$ supplemented with the appropriate $b_\mu$ dependent terms such that all the $b_\mu$ terms drop out on the right hand side of \eqref{eq:covder1v2}--\eqref{eq:covder4v2} when rewriting it in terms of the connections $\Gamma^{\rho}_{\mu\nu}$, $\Omega_\mu{}^a$ and $\Omega_\mu{}^{ab}$. Solving the vielbein postulates \eqref{eq:VP1}--\eqref{eq:VP4} for $\omega_\mu{}^a$ and $\omega_\mu{}^{ab}$ in terms of $b_\mu$ and $\tilde\Gamma^{\rho}_{\mu\nu}$ we can write the result as
\begin{eqnarray}
\omega_\mu{}^a & = & -\frac{1}{2} v^\nu\left(\partial_\mu e_\nu{}^a-\partial_\nu e_\mu{}^a\right)+\frac{1}{2}e_\mu{}^c v^\nu e^{\rho a}\left(\partial_\nu e_{\rho c}-\partial_\rho e_{\nu c}\right)\nonumber\\
&&-\frac{1}{2}e^{\sigma a}\left(\partial_\sigma M_\mu-\partial_\mu M_\sigma-(2-z)(b_\sigma M_\mu-b_\mu M_\sigma)\right)\nonumber\\
&&+\frac{1}{2}\tau_\mu e^{\sigma a}v^\nu\left(\partial_\sigma M_\nu-\partial_\nu M_\sigma-(2-z)(b_\sigma M_\nu-b_\nu M_\sigma)\right)\nonumber\\
&&-e_\mu{}^a v^\nu b_\nu+v^\nu e^{\sigma a}\left(M_\sigma T_{\mu\nu}-M_\mu T_{\nu\sigma}-M_\nu T_{\mu\sigma}\right) \,,\label{eq:tildeomegaboost}\\
\omega_\mu{}^a{}_b & = & \frac{1}{2}e^\nu{}_b\left(\partial_\mu e_\nu{}^a-\partial_\nu e_\mu{}^a\right)-\frac{1}{2}e_\mu{}^c e^{\rho a}e^\nu{}_b\left(\partial_\nu e_{\rho c}-\partial_\rho e_{\nu c}\right)-\frac{1}{2}e^{\rho a}\left(\partial_\mu e_{\rho b}-\partial_\rho e_{\mu b}\right)\nonumber\\
&&-\frac{1}{2}\tau_\mu e^{\sigma a}e^\nu{}_b\left(\partial_\sigma M_\nu-\partial_\nu M_\sigma-(2-z)(b_\sigma M_\nu-b_\nu M_\sigma)\right)+e_\mu{}^a e^\nu{}_b b_\nu-e_{\mu b}e^{\nu a}b_\nu\nonumber\\
&&-e^{\sigma a}e^\nu{}_b\left(M_\sigma T_{\mu\nu}-M_\mu T_{\nu\sigma}-M_\nu T_{\mu\sigma}\right)\,,\label{eq:tildeomegarotation}
\end{eqnarray}
where $T_{\mu\nu}$ is the twist tensor defined as
\begin{equation}
T_{\mu\nu}=\frac{1}{2}\bar h_{\mu\rho}\bar h_{\nu\sigma}h^{\rho\lambda}h^{\sigma\kappa}\left(\partial_\lambda\tau_\kappa-\partial_\kappa\tau_\lambda\right)\,.
\end{equation}
Note that for TTNC geometry we have that $T_{\mu\nu}=0$ due to \eqref{hypsurforth1} (which for general $z$ is just $\partial_{[\mu} \tau_{\nu]} = z b_{[\mu} \tau_{\nu]}$) and that in this case the expressions \eqref{eq:tildeomegaboost} and \eqref{eq:tildeomegarotation} agree with \eqref{eq:omega-ab} and \eqref{eq:omega-a} (where of course one must replace $m_\mu$ by $M_\mu$ as explained in section \ref{sec:Stueckelbergscalarzneq2}). In deriving these results it is useful to use the fact that one can write \eqref{eq:TNCb} equivalently as
\begin{equation}\label{eq:def-b}
\partial_\mu \tau_\nu-\partial_\nu\tau_\mu=-z\left(\tau_\mu b_\nu-\tau_\nu b_\mu\right)+2T_{\mu\nu}\,,
\end{equation}
which is the same statement as the vanishing of the antisymmetric part of \eqref{eq:covder1v2}.

Combining \eqref{eq:def-b} with \eqref{eq:tildeGammaTNC} we see that for TNC geometry the connection $\tilde\Gamma^\rho_{\mu\nu}$ becomes torsionful with torsion given by
\begin{equation}\label{eq:TNCtorsion}
\tilde\Gamma^{\rho}_{[\mu\nu]}= -\hat v^\rho T_{\mu\nu}\,,
\end{equation}
and that this becomes torsionless if and only if we are dealing with a TTNC geometry.

\subsection{The central charge gauge connection $m_\mu$}

The introduction of the $b_\mu$ field, via \eqref{eq:TNCb}, also allows us to define the central charge gauge connection $m_\mu$. The definition of $m_\mu$ follows from the observation made in section \ref{sec:Stueckelbergscalarzneq2}, equation \eqref{eq:defM}, that one can view $M_\mu$ as the covariant derivative of $\chi$ by writing
\begin{equation}
M_\mu=-\mathcal{D}_\mu\chi=m_\mu+(2-z)\chi b_\mu-\partial_\mu\chi\,.
\end{equation}
Here we use this as the definition of $m_\mu$. It follows that  $m_\mu$ must transform as in \eqref{eq:gaugetrafom}.

Since $M_\mu$ also transforms under boosts the covariant derivative acting on $M_\mu$ is
\begin{equation}
\mathcal{D}_\mu M_\nu=\partial_\mu M_\nu-\tilde\Gamma^{\rho}_{\mu\nu}M_\rho-(2-z)b_\mu M_\nu-\omega_\mu{}^a e_{\nu a}\,.
\end{equation}
It follows that we obtain
\begin{equation}\label{eq:constraintchi2}
\left[\mathcal{D}_\mu\,,\mathcal{D}_\nu\right]\chi =-2\mathcal{D}_{[\mu} M_{\nu]}=4 T_{\mu\nu}\tilde\Phi\,,
\end{equation}
where $\tilde\Phi$ is given in \eqref{eq:tildePhi}. One can interpret this as the TNC generalization of the curvature constraint \eqref{eq:newcurvconstraint}. Likewise equation \eqref{eq:def-b} can be viewed as the TNC generalization of the curvature constraint $R_{\mu\nu}(H)=0$ used in the gauging of the Schr\"odinger algebra. Finally the TNC analogue of the constraint $R_{\mu\nu}{}^a(P)=0$ is the vanishing of the antisymmetric part of \eqref{eq:covder2v2}.

For later purposes we mention that from the vielbein postulates \eqref{eq:VP1}--\eqref{eq:VP4} with the covariant derivatives written as in \eqref{eq:covder1v2}--\eqref{eq:covder4v2} it follows that
\begin{equation}\label{eq:VPhate}
\mathcal{D}_\mu\hat e_\nu{}^a=-\tau_\nu\mathcal{D}_\mu M^a
\end{equation}
where $\hat e_\nu{}^a$ is given in \eqref{eq:hate} and where $\mathcal{D}_\mu M^a$ is defined as
\begin{equation}\label{eq:covderMa}
\mathcal{D}_\mu M^a=\partial_\mu M^a-\omega_\mu{}^a-\omega_\mu{}^a{}_bM^b-(1-z)b_\mu M^a\,,
\end{equation}
so that $\mathcal{D}_\mu M^a$ is boost invariant.

\subsection{Special conformal transformations}

In section \ref{subsec:enhancement} we showed that for TTNC with general $z$ there is an extra symmetry that allows us to remove the temporal component of the $b_\mu$ connection. We will show in this and the next subsection that this symmetry also exists for TNC geometry. In other words we will show that there is a new symmetry of the form \begin{equation}\label{eq:Ktrafo}
\delta_K b_\mu=\Lambda_K\tau_\mu\,.
\end{equation}
What this means is that the field $\hat v^\mu b_\mu$ in \eqref{eq:TNCb} can be gauged away using this symmetry. This symmetry must leave invariant the fields $M_\mu$, $\tau_\mu$ and $e_\mu{}^a$ since we do not see it in the holographic context of \cite{Hartong:2014pma,Hartong:2014oma}. Since $\delta_K M_\mu=\delta_K \chi=0$ we need that
\begin{equation}
\delta_K m_\mu=-(2-z)\Lambda_K\chi\tau_\mu\,.
\end{equation}
Apart from $b_\mu$ and $f_\mu$ the only other field that transforms under $\delta_K$ is $\omega_\mu{}^a$ because it depends on $v^\rho b_\rho$. Using \eqref{eq:tildeomegaboost} we get
\begin{equation}
\delta_K\tilde\omega_\mu{}^a=\Lambda_K\left(\hat e_\mu{}^a+(z-1)\tau_\mu M^a\right)\,.
\end{equation}
The discussion is very analogous to the discussion of conformal symmetries in the TTNC case of section \ref{subsec:enhancement} so we shall be brief and merely highlight the new ingredients.

We introduce a curvature for $b_\mu$ denoted as usual by $R_{\mu\nu}(D)$ which is given by
\begin{equation}
R_{\mu\nu}(D)=\partial_\mu b_\nu-\partial_\nu b_\mu-f_\mu\tau_\nu+f_\nu\tau_\mu\,.
\end{equation}
We next demand that it is invariant under all transformations except under the $K$ transformation in which case it transforms like \eqref{eq:trafonewDcurv}. This tells us that $f_\mu$ must transform as
\begin{equation}\label{eq:trafo-f2}
\delta f_\mu=\partial_\mu\Lambda_K-z\Lambda_D f_\mu+z\Lambda_K b_\mu\,.
\end{equation}
Since now we have the constraint \eqref{eq:def-b} it follows that
\begin{equation}
\delta R_{\mu\nu}(D)=2\Lambda_K T_{\mu\nu}\,.
\end{equation}
Hence we need a constraint of the form
\begin{equation}\label{eq:constraintf}
R_{\mu\nu}(D)=\partial_\mu b_\nu-\partial_\nu b_\mu-f_\mu\tau_\nu+f_\nu\tau_\mu=2T_{(b)\mu\nu}\,,
\end{equation}
where $T_{(b)\mu\nu}$ is an object that is invariant under all transformations except under the $K$ transformation in which case it goes like $\delta_K T_{(b)\mu\nu}=2T_{\mu\nu}$. This object $T_{(b)\mu\nu}$ is given by
\begin{equation}
T_{(b)\mu\nu}=\frac{1}{2}\bar h_{\mu\rho}\bar h_{\nu\sigma}h^{\rho\kappa}h^{\sigma\lambda}\left(\partial_\kappa b_\lambda-\partial_\lambda b_\kappa\right)\,.
\end{equation}
One can show that for TTNC we have $\Omega_{(b)\mu\nu}=0$ as it should be.

\subsection{The connection $f_\mu$}

To realize $f_\mu$ as a dependent gauge connection we start by using \eqref{eq:constraintf} from which we find
\begin{equation}\label{eq:dependentfpart1}
f_\mu=\hat v^\rho\left(\partial_\rho b_\mu-\partial_\mu b_\rho\right)-\hat v^\rho f_\rho\tau_\mu\,,
\end{equation}
where in obtaining this expression we used the fact that $\hat v^\mu T_{(b)\mu\nu}=0$. We thus need to find an expression for $\hat v^\rho f_\rho$ which transforms as
\begin{equation}\label{eq:whatweneed}
\delta\left(\hat v^\mu f_\mu\right)=\hat v^\mu\partial_\mu\Lambda_K-2z\Lambda_D\hat v^\mu f_\mu+z\Lambda_K\hat v^\mu b_\mu\,,
\end{equation}
in order that $f_\mu$ transforms as \eqref{eq:trafo-f2}. Of course we know that $\hat v^\mu f_\mu=v^\mu f_\mu-h^{\mu\nu}M_\nu f_\mu$ and we already have an expression for $h^{\mu\nu}M_\nu f_\mu$ due to \eqref{eq:dependentfpart1}. Working out its transformation we find
\begin{equation}\label{eq:whatweknow}
\delta\left(h^{\mu\nu}M_\nu f_\mu\right)=h^{\mu\nu}M_\nu \partial_\mu\Lambda_K+\lambda^a e^\mu{}_a f_\mu-2z\Lambda_Dh^{\mu\nu}M_\nu f_\mu+z\Lambda_K h^{\mu\nu}M_\nu b_\mu\,.
\end{equation}
Hence subtracting \eqref{eq:whatweknow} from \eqref{eq:whatweneed} we learn that we need to find an expression for $v^\mu f_\mu$ that transforms as 
\begin{equation}\label{eq:whatweneed2}
\delta\left(v^\mu f_\mu\right)=\lambda^a e^\mu{}_a f_\mu+v^\mu\partial_\mu\Lambda_K-2z\Lambda_D v^\mu f_\mu+z\Lambda_K v^\mu b_\mu\,.
\end{equation}

We will construct $v^\mu f_\mu$ by making an ansatz further below. The most difficult aspect is to get the transformation with respect to boosts to work out. For this purpose it will prove very convenient to establish a Bianchi identity which we now derive. Acting with a covariant derivative on \eqref{eq:covderMa}, given by
\begin{equation}
\mathcal{D}_\mu\mathcal{D}_\nu M^a=\partial_\mu\left(\mathcal{D}_\nu M^a\right)-\tilde\Gamma^{\rho}_{\mu\nu}\mathcal{D}_\rho M^a-(1-z)b_\mu\mathcal{D}_\nu M^a-\tilde\omega_\mu{}^a{}_b\mathcal{D}_\nu M^b+f_\mu \hat e_\nu{}^a\,,
\end{equation}
we see that
\begin{equation}\label{eq:Gcurvature}
\left[\mathcal{D}_\mu\,,\mathcal{D}_\nu\right] M^a=-2\tilde\Gamma^{\rho}_{[\mu\nu]}\mathcal{D}_\rho M^a-R_{\mu\nu}{}^{ab}(J)M_b-(1-z)R_{\mu\nu}(D)M^a-R_{\mu\nu}{}^a(G)\,,
\end{equation}
where the curvatures are the same as those given at the end of section \ref{subsec:enhancement}. Multiplying \eqref{eq:Gcurvature} by $-\tau_\rho$ and antisymmetrizing over all indices using \eqref{eq:VPhate} and \eqref{eq:hate} gives
\begin{eqnarray}
2\mathcal{D}_{[\mu}\mathcal{D}_{\nu}\hat e_{\rho]}{}^a & = & -2\tilde\Gamma^{\sigma}_{[\mu\nu}\mathcal{D}_{\vert\sigma\vert} \hat e_{\rho]}{}^a+R_{[\mu\nu}{}^{ab}(J)(e_{\rho]b}-\hat e_{\rho]b})+(1-z)R_{[\mu\nu}(D)(e_{\rho]}{}^a-\hat e_{\rho]}{}^a)\nonumber\\
&&+R_{[\mu\nu}{}^a(G)\tau_{\rho]}\,.
\end{eqnarray}
Next we use the identity
\begin{equation}
2\mathcal{D}_{[\mu}\mathcal{D}_{\nu}\hat e_{\rho]}{}^a=-2\tilde\Gamma^{\sigma}_{[\mu\nu}\mathcal{D}_{\vert\sigma\vert} \hat e_{\rho]}{}^a+2\tilde\Gamma^{\sigma}_{[\mu\nu}\mathcal{D}_{\rho]} \hat e_{\sigma}{}^a-R_{[\mu\nu}{}^{ab}(J)\hat e_{\rho]b}-R_{[\mu\nu}(D)\hat e_{\rho]}{}^a
\end{equation}
to derive the Bianchi identity
\begin{equation}\label{eq:BInew}
0 = 2\Omega_{[\mu\nu}\mathcal{D}_{\rho]}M^a+2\Omega_{(b)[\mu\nu}e_{\rho]}{}^a-2zM^a\Omega_{(b)[\mu\nu}\tau_{\rho]}+R_{[\mu\nu}{}^{ab}(J)e_{\rho]b}+R_{[\mu\nu}{}^a(G)\tau_{\rho]}\,.
\end{equation}
By contracting this with $v^\mu e^\nu{}_c e^\rho{}_a$ we find
\begin{eqnarray}
R_{ca}{}^a(G) + v^\mu R_{\mu a}{}^a{}_c(J) & =  & 2M^b\Omega_{bc}e^\nu{}_a\mathcal{D}_\nu M^a-2M^b\Omega_{ba}e^\nu{}_c\mathcal{D}_\nu M^a+2\Omega_{ca}v^\mu\mathcal{D}_\mu M^a\nonumber\\
&&+2(d-z-1)M^b\Omega_{(b)bc}\,,\label{eq:BI1}
\end{eqnarray}
and by contracting \eqref{eq:BInew} with $e^\mu{}_b e^\nu{}_a e^\rho{}_c$ we obtain
\begin{eqnarray}
R_{ba}{}^a{}_c(J)-R_{ca}{}^a{}_b(J) & = & -2\Omega_{ba}e^\nu{}_c\mathcal{D}_\nu M^a+2\Omega_{ca}e^\nu{}_b\mathcal{D}_\nu M^a+2\Omega_{bc}e^\nu{}_a\mathcal{D}_\nu M^a\nonumber\\
&& -2(2-d)\Omega_{(b)bc}\,.\label{eq:BI2}
\end{eqnarray}

We make the following ansatz for $v^\mu f_\mu$
\begin{eqnarray}
v^\mu f_\mu & = & F+\frac{1}{d}v^\mu e^\nu{}_a\left(2\partial_{[\mu}\omega_{\nu]}{}^a-2\omega_{[\mu}{}^{ab}\omega_{\nu]b}-2(1-z)b_{[\mu}\omega_{\nu]}{}^a-2(z-2)f_{[\mu}\tau_{\nu]}M^a\right)\,,
\end{eqnarray}
where $F$ needs to be determined. The term in parenthesis makes up a large part of the curvature $R_{\mu\nu}{}^a(G)$ given in \eqref{eq:newGcurv}. Note that the right hand side does not contain $v^\mu f_\mu$ but only $e^\mu{}_a f_\mu$ which we already know. This transforms under boosts as\footnote{The transformation properties of the connections $\omega_{\mu}{}^{ab}$ and $\omega_{\mu}{}^b$ follow readily from the vielbein postulates \eqref{eq:VP1}--\eqref{eq:VP4} written using \eqref{eq:covder1v2}--\eqref{eq:covder4v2} and they are the same as for TTNC geometry.}
\begin{equation}
\delta_G \left(v^\mu f_\mu\right)=\delta_G F+\lambda^a e^\mu{}_a f_\mu+\frac{2}{d}(z-1)\lambda^cv^\mu\Omega_{(b)\mu c}+\frac{1}{d}\lambda^c\left[R_{ca}{}^a(G)-v^\mu R_{\mu a}{}^a{}_c(J)\right]\,.
\end{equation}
In order to cancel the curvature terms upon use of the Bianchi identities \eqref{eq:BI1} and \eqref{eq:BI2} we take for $F$
\begin{equation}
F=\frac{2}{d}\hat v^\mu M^c R_{\mu a}{}^a{}_c(J)+\frac{1}{d}M^b M^c R_{ba}{}^a{}_c(J)+\frac{2}{d}\Omega_{ac}M^c\hat v^\mu\mathcal{D}_\mu M^a+\tilde F\,.
\end{equation}
This leads to 
\begin{equation}
\delta_G \left(v^\mu f_\mu\right)=\delta_G\tilde F+\lambda^a e^\mu{}_a f_\mu\,.
\end{equation}
We will take $\delta_G\tilde F=0$ so that we obtain a boost invariant expression for $f_\mu$. We still need to ensure that $f_\mu$ transforms as in \eqref{eq:trafo-f2} with respect to dilatations and special conformal transformations. It is straightforward to check that under dilatations we have
\begin{equation}
\delta_D\left(v^\mu f_\mu\right)=-2z\Lambda_D v^\mu f_\mu\,,
\end{equation}
provided we take $\delta_D\tilde F=-2z\Lambda_D\tilde F$. Finally under special conformal transformations we have
\begin{equation}
\delta_K\left(v^\mu f_\mu\right)=\delta_K\tilde F+v^\mu\partial_\mu\Lambda_K+z\Lambda_K v^\mu b_\mu+\frac{1}{d}(z-2)\Lambda_K e^\mu{}_a\mathcal{D}_\mu M^a\,.
\end{equation}
The last term can be cancelled by taking
\begin{equation}
\tilde F=\frac{1}{2d^2}(z-2)\left(e^\mu{}_a\mathcal{D}_\mu M^a\right)^2\,,
\end{equation}
where using \eqref{eq:covderMa} and the expression for $\omega_\mu{}^a$ and $\omega_\mu{}^{ab}$ given in \eqref{eq:tildeomegaboost} and \eqref{eq:tildeomegarotation}, respectively, we have
\begin{equation}
e^\mu{}_a\mathcal{D}_\mu M^a=-\frac{1}{e}\partial_\mu\left(e\hat v^\mu\right)+d\hat v^\mu b_\mu\,,
\end{equation}
with $e$ the determinant of the matrix formed by $(\tau_\mu\,, e_\mu{}^a)$. One sees that $e^\mu{}_a\mathcal{D}_\mu M^a$ is boost invariant and has dilatation weight $-z$ so that $\tilde F$ obeys the conditions $\delta_G\tilde F=0$ and $\delta_D\tilde F=-2z\Lambda_D\tilde F$. The final expression for $v^\mu f_\mu$ is thus
\begin{eqnarray}
v^\mu f_\mu & = & \frac{1}{d}v^\mu e^\nu{}_a\left(2\partial_{[\mu}\tilde\omega_{\nu]}{}^a-2\tilde\omega_{[\mu}{}^{ab}\tilde\omega_{\nu]b}-2(1-z)b_{[\mu}\tilde\omega_{\nu]}{}^a-2(z-2)f_{[\mu}\tau_{\nu]}M^a\right)\nonumber\\
&&+\frac{2}{d}\hat v^\mu M^c R_{\mu a}{}^a{}_c(J)+\frac{1}{d}M^b M^c R_{ba}{}^a{}_c(J)+\frac{2}{d}\Omega_{ac}M^c\hat v^\mu\mathcal{D}_\mu M^a\nonumber\\
&&+\frac{1}{2d^2}(z-2)\left(e^\mu{}_a\mathcal{D}_\mu M^a\right)^2\,.\label{eq:vf}
\end{eqnarray}
We thus see that $f_\mu$ is a completely dependent gauge connection given by \eqref{eq:dependentfpart1} and \eqref{eq:vf}.

\subsubsection{From TTNC to TNC}

This completes the description of TNC geometry. We conclude that TNC geometry is an extension of TTNC obtained by relaxing the TTNC curvature constraints to
\begin{eqnarray}
R_{\mu\nu}(H) & = & 2T_{\mu\nu}\,,\\
R_{\mu\nu}(D) & = & 2T_{(b)\mu\nu}\,,\label{eq:curvconsD}\\
R_{\mu\nu}(N) & = & 2T_{\mu\nu}v^\rho M_\rho+2(z-2)\chi T_{(b)\mu\nu}\,,\\
R_{\mu\nu}{}^a(P) & = & 2T_{\mu\nu}M^a\,,\label{eq:TNCPconstraint}
\end{eqnarray}
where the curvatures $R_{\mu\nu}(D)$ and $R_{\mu\nu}(N)$ are given at the end of section \ref{subsec:enhancement} and $R_{\mu\nu}(H)$ and $R_{\mu\nu}{}^a(P)$ can be found in section \ref{subsubsec:gaugetrafos-constraints-zneq2}. The constraint from which $v^\mu f_\mu$ can be obtained is
\begin{eqnarray}
0 & = & \hat v^\mu\left( R_{\mu a}{}^a(G)+2 M^c R_{\mu a}{}^a{}_c(J)\right)+M^b\left(R_{ba}{}^a(G)+M^c R_{ba}{}^a{}_c(J)\right) \nonumber\\
&&+2\Omega_{ac}M^c\hat v^\mu\mathcal{D}_\mu M^a+\frac{1}{2d}(z-2)\left(e^\mu{}_a\mathcal{D}_\mu M^a\right)^2\,.
\end{eqnarray}
To prove \eqref{eq:TNCPconstraint} we used the antisymmetric part of $\mathcal{D}_{[\mu}e_{\nu]}{}^a=0$ together with \eqref{eq:TNCtorsion}. To derive the curvature constraint for $R_{\mu\nu}(M)$ we used \eqref{eq:constraintchi2} together with
\begin{equation}
\left[\mathcal{D}_\mu\,,\mathcal{D}_\nu\right]\chi =-2\tilde\Gamma^{\rho}_{[\mu\nu]}\mathcal{D}_\rho\chi-R_{\mu\nu}(N)-(2-z)\chi R_{\mu\nu}(D)
\end{equation}
and \eqref{eq:TNCtorsion}.

\section{Conclusions} \label{sec:conclusions}

Torsional Newton--Cartan geometry is expected to play an important role in Lifshitz holography, where it can serve as a guiding principle to construct  precise holographic dictionaries. In this paper, we have constructed a  vielbein formulation for generic torsional Newton--Cartan geometry, putting special emphasis on the Schr\"odinger-type local symmetries that are needed in the construction. Our approach has at first been to perform a gauging of the Schr\"odinger algebra. In this procedure gauge fields are introduced for all generators of the Schr\"odinger algebra, whose transformation rules and covariant curvatures are determined by the structure constants of the algebra. One also imposes curvature constraints to make certain gauge fields dependent on the remaining ones and to identify diffeomorphisms and local space-time translations. We have shown that in this way, one can indeed define a vielbein formalism for a specific kind of torsional Newton--Cartan geometry, so-called twistless torsional Newton--Cartan geometry. 

For applications to Lifshitz holography, a more general procedure is however required. Indeed, Lifshitz holography allows for more general kinds of torsional Newton--Cartan geometries than the twistless torsional ones. Furthermore, the central charge of the Schr\"odinger algebra is promoted to a St\"uckelberg symmetry, in the sense that it is accompanied by an extra St\"uckelberg scalar for the central charge. In contrast, in the ordinary gauging of the Schr\"odinger algebra, this St\"uckelberg scalar is not present. With applications to Lifshitz holography in mind, we have therefore shown how (twistless) torsional Newton--Cartan geometry, in the presence of the St\"uckelberg scalar can be defined via a procedure, inspired by the gauging of the Schr\"odinger algebra. In particular, we have shown how one can introduce gauge fields, associated transformation rules and covariant curvatures for Schr\"odinger-type symmetries, in the presence of the St\"uckelberg scalar. We have argued how curvature constraints turn some gauge fields into dependent ones and that this procedure indeed leads to (twistless) torsional Newton--Cartan geometry, as it appears as boundary geometry in Lifshitz holography.

The appearance of local Schr\"odinger-type symmetries in Lifshitz holography might perhaps seem odd. However, as we hope to have elucidated in this paper, from the point of view of the torsional Newton--Cartan boundary geometry, it is rather natural. Indeed, the description of this geometry requires the presence of an extra vector field, apart from the temporal and spatial vielbeine, that can be associated to a central charge transformation. The local boundary symmetries should therefore not only include scale transformations, but should also include such a central charge transformation. This naturally leads one to look at Schr\"odinger-type symmetries. In this paper, we have clarified how these symmetries are precisely connected to torsional Newton--Cartan geometry. The connection between Schr\"odinger-type symmetries and torsional Newton--Cartan geometry, that we have studied in this paper, has implications for Lifshitz holography. Implications for the dual field theories that Lifshitz holography attempts to describe have been explored in \cite{Hartong:2014pma,Hartong:2014oma}. 

Although technical in nature, we hope that this paper clarifies a number of issues, regarding the local symmetries and geometries that are realized in Lifshitz holography. Given how symmetries and their potential geometric realization have always played an important role in the construction of effective field theories, we expect our work to be of use in the more general context of the study of non-relativistic field theories.

In this work we have described TNC geometry in terms of the fields $\tau_\mu$, $e_\mu{}^a$, $m_\mu$ and $\chi$. This is naturally suggested by the holographic setup in which the bulk geometry is described by Einstein gravity coupled to a massive vector field and possibly a dilaton. There are however other setups leading to Lifshitz space-times, such as the Einstein--Maxwell-dilaton model with a logarithmically running dilaton \cite{Taylor:2008tg}. In this case we have Einstein gravity coupled to a Maxwell bulk field and a dilaton. It would be interesting to see what kind of boundary geometry we would get in this case and if it is the same or different from what we found here. More generally one can add another exponent \cite{Gouteraux:2012yr,Gath:2012pg} related to the logarithmic running of the dilaton on top of the critical exponent $z$ and still have a Lifshitz geometry. It would be interesting to study the role of this exponent from the point of view of the boundary geometry. One can also consider the use of Horava--Lifshitz gravity in the bulk \cite{Horava:2009uw} which admits Lifshitz space-times as a vacuum solution \cite{Griffin:2012qx}. It would be interesting to see what the dual geometry is, how it comes about from the bulk perspective (see \cite{Wu:2014dha} for work in this direction), whether Schr\"odinger symmetries play a role, in particular in relation to particle number, and whether there is again a $\chi$ field or whether this gets replaced by something else.

\section*{Acknowledgements}

We would like to thank Jay Armas, Arjun Bagchi, Matthias Blau, Jan de Boer, Kristan Jensen, Elias Kiritsis, and especially Niels Obers for many very useful discussions. We are especially grateful to Joaquim Gomis for collaboration in the initial stages of this work and for sharing many valuable insights. The work of J.H. is supported in part by the Danish National Research Foundation project ``Black holes and their role in quantum gravity". J.H. wishes to thank the University of Groningen and CERN for their hospitality and financial support. The work of J.R. is supported by the START project Y 435-N16 of the Austrian Science Fund (FWF). E.B and J.R. would like to thank the organizers of the ``Quantum Gravity, Black Holes and Strings" program at the KITPC-Beijing for financial support and hospitality when part of this work was done.


\providecommand{\href}[2]{#2}\begingroup\raggedright\endgroup

\end{document}